\documentclass[11pt,a4paper]{article}                           

\usepackage{tikz}
\usepackage{tikz-3dplot}
\usetikzlibrary{calc}
\usetikzlibrary{decorations} %
\usepackage[english]{babel}
\usepackage{amsmath}
\usepackage{amssymb}
\usepackage{graphicx}
\usepackage{wrapfig}
\usepackage{hhline}
\usepackage[bf]{caption}
\usepackage{cite}
\usepackage[vcentermath]{youngtab}
\usepackage[pagebackref]{hyperref} 
\usepackage{xspace}

\usepackage{tikz-cd} 
 
\makeatletter

\renewcommand*{\backref}[1]{}
\renewcommand*{\backrefalt}[4]{({%
    \ifcase #1 Not cited.%
          \or Page~#2.%
          \else Pages #2.
    \fi%
    })} 
 
\usepackage{multirow}
\usepackage{longtable,array}

\makeatletter
\def\nobreakhline{%
  \noalign{\ifnum0=`}\fi
    \penalty\@M
    \futurelet\@let@token\LT@@nobreakhline}
\def\LT@@nobreakhline{%
  \ifx\@let@token\hline
    \global\let\@gtempa\@gobble
    \gdef\LT@sep{\penalty\@M\vskip\doublerulesep}% <-- change here
  \else
    \global\let\@gtempa\@empty
    \gdef\LT@sep{\penalty\@M\vskip-\arrayrulewidth}% <-- change here
  \fi
  \ifnum0=`{\fi}%
  \multispan\LT@cols
     \unskip\leaders\hrule\@height\arrayrulewidth\hfill\cr
  \noalign{\LT@sep}%
  \multispan\LT@cols
     \unskip\leaders\hrule\@height\arrayrulewidth\hfill\cr
  \noalign{\penalty\@M}%
  \@gtempa}
\makeatother
\numberwithin{equation}{section}
\numberwithin{table}{section}

\newcommand{\KB}{\overline{K}_B}
\newcommand{\bbZ}{{\mathbb{Z}}}
\newcommand{\bbP}{{\mathbb{P}}}
\newcommand{\cO}{{\cal O}}
\newcommand{\cA}{{\cal A}}

\newcommand{\one}[0]{\ensuremath{\mathbf{1} }\xspace}
\newcommand{\two}[0]{\ensuremath{\mathbf{2} }\xspace}

\newcommand{\three}[0]{\ensuremath{\mathbf{3} }\xspace}

\newcommand{\executeiffilenewer}[3]{%
 \ifnum\pdfstrcmp{\pdffilemoddate{#1}}%
 {\pdffilemoddate{#2}}>0%
 {\immediate\write18{#3}}\fi%
}

\makeatother
\hypersetup{
    pdftitle={An F-theory Realization of the Chiral MSSM with Z2-Parity},
    pdfauthor={Mirjam Cvetic, Ling Lin, Muyang Liu, Paul-Konstantin Oehlmann},
    pdfsubject={F-theory compactifications, Standard Model, Matter parity, G4-flux}
}

\begin{document}

\baselineskip=15pt
\begin{titlepage}
 
{\hfill \texttt{UPR-1292-T}}

\vspace*{1.5cm} 
\begin{center}
 {\LARGE  An F-theory Realization of the Chiral MSSM with $\bbZ_2$-Parity}\\

 \vspace*{1.8cm}
 {Mirjam Cveti\v{c}$\,^{1,2}$, Ling Lin$\,^1$, Muyang Liu$\,^1$, Paul-Konstantin Oehlmann$\,^3$}\\

 \vspace*{1.2cm} 

{\it $^1$ Department of Physics and Astronomy, University of Pennsylvania,\\
 Philadelphia, USA}\\

 \bigskip
{\it $^2$ Center for Applied Mathematics and Theoretical Physics, \\
 University of Maribor, Maribor, Slovenia}\\

\bigskip

{\it $^3$ Physics Department, Robeson Hall, Virginia Tech,\\
 Blacksburg, USA}\\

 \bigskip

\vspace*{0.8cm}
\end{center}
\vspace*{.5cm}

\begin{abstract}
\noindent
Using F-theory we construct 4D ${\cal N}=1$ SUGRA theories with the Standard Model gauge group, three chiral generations, and matter parity in order to forbid all dimension four baryon and lepton number violating operators. The underlying geometries are derived by constructing smooth genus-one fibered Calabi--Yau fourfolds using toric tops that have a Jacobian fibration with rank one Mordell--Weil group and $SU(3) \times SU(2) $ singularities. 
The necessary gauge backgrounds on the smooth fourfolds are shown to be fully compatible with the quantization condition, including positive integer D3-tadpoles.
This construction realizes for the first time a consistent UV completion of an MSSM-like model with matter parity in F-theory.
Moreover our construction is general enough to also exhibit other relevant $\mathbb{Z}_2$ charge extensions of the MSSM such as lepton and baryon parity. 
Such models however are rendered inconsistent by non-integer fluxes, which are necessary for producing the exact MSSM chiral spectrum.
These inconsistencies turn out to be intimately related to field theory considerations regarding a UV-embedding of the $\bbZ_2$ into a $U(1)$ and the resulting discrete anomalies.
\end{abstract}

\end{titlepage}
\clearpage
\setcounter{footnote}{0}
\setcounter{tocdepth}{2}
\tableofcontents
\clearpage

\baselineskip=14pt
\parskip 5pt plus 1pt 

\interfootnotelinepenalty=10000

%
%%%%%%%%%%%%%%%%%%%%%%%%%%%%%%%%%%%%%%%%%%%%%%%%
%%%%%%%%%%%%%%%%%%%%%%%%%%%%%%%%%%%%%%%%%%%%%%%%
%%%%%%%%%%%%%%%%%%%%%%%%%%%%%%%%%%%%%%%%%%%%%%%%
%%%%%%%%%%%%%%%%%%%%%%%%%%%%%%%%%%%%%%%%%%%%%%%%
%%%%%%%%%%%%%%%%%%%%%%%%%%%%%%%%%%%%%%%%%%%%%%%%
%%%%%%%%%%%%%%%%%%%%%%%%%%%%%%%%%%%%%%%%%%%%%%%%
%%%%%%%%%%%%%%%%%%%%%%%%%%%%%%%%%%%%%%%%%%%%%%%%
%%%%%%%%%%%%%%%%%%%%%%%%%%%%%%%%%%%%%%%%%%%%%%%%
%
%\begin{abstract}
%
%
% 
%
%
%
%\end{abstract}
% 
%\vspace*{\fill}
%
%
%\thispagestyle{empty}
%\clearpage
%\setcounter{page}{1}

%%%%%%%%%%%%%%%%%%%%%%%%%%%%%%%%%%%%%%%%%%%%%%%
%%%%%%%%%%%%%%%%%%%%%%%%%%%%%%%%%%%%%%%%%%%%%%%
%%%%%%%%%%%                 %%%%%%%%%%%%%%%%%%%
%%%%%%%%%%%  DOCUMENT BODY  %%%%%%%%%%%%%%%%%%%
%%%%%%%%%%%                 %%%%%%%%%%%%%%%%%%%
%%%%%%%%%%%%%%%%%%%%%%%%%%%%%%%%%%%%%%%%%%%%%%%
%%%%%%%%%%%%%%%%%%%%%%%%%%%%%%%%%%%%%%%%%%%%%%%
%%%%%%%%%%%%%%%%%%%%%%%%%%%%%%%%%%%%%%%%%%%%%%%

%\newpage

%\tableofcontents
%%%%%%%%%%%%%%%
\section{Introduction}
\label{sec:one}
One of the major goals of string theory is to provide a possible framework to UV complete the Standard Model of particle physics together with gravity. 
The web of string theories, connected by various dualities, exhibits a rich landscape of possibilities, and each corner provides an interesting starting point towards this goal, coming with its own benefits and challenges.
In particular, F-theory \cite{Vafa:1996xn} provides a very flexible tool to cover the to date largest set of consistent vacua \cite{Taylor:2015ppa,Taylor:2015xtz,Halverson:2016tve,Halverson:2017ffz} from string theory by using a non-peturbative extension of Type IIB strings (see \cite{Taylor:2011wt,Weigand:2010wm,Maharana:2012tu,Weigand:2018rez} for some reviews). 

This flexibility lies in the systematic engineering of gauge theories coupled to gravity using powerful tools of algebraic geometry, that combines the strength of various perturbative string theories.
In its early days, F-theory has been employed to engineer the whole Standard Model through a unified gauge group $SU(5)$ on a single divisor \cite{Donagi:2008ca,Beasley:2008dc,Beasley:2008kw}, which then has to be broken to the Minimal Supersymmetric Standard Model (MSSM) by a flux in the hypercharge Cartan subgroup \cite{Donagi:2008kj}. The localization of the $SU(5)$ made a simple local treatment of the compactification space possible, although it was required to enhance these models by additional Abelian symmetries to control proton decay and Yukawa textures \cite{Dudas:2009hu}. 
The global realization of Abelian gauge symmetries and its connection to the Mordell--Weil group of the fibration, although already pointed out earlier \cite{Morrison:1996pp}, was only further explored later on in \cite{Grimm:2010ez, Morrison:2012ei,Cvetic:2012xn}. 
These developments kicked off the construction of globally consistent realizations of GUTs together with $U(1)$ symmetries \cite{Mayrhofer:2012zy, Braun:2013nqa, Borchmann:2013hta,Cvetic:2013uta,Krippendorf:2014xba, Lawrie:2015hia, Krippendorf:2015kta}. 
Nevertheless, these construction were still relying on a GUT breaking mechanism via hypercharge flux \cite{Braun:2014pva} which is often technically hard to implement or might lead to vector-like exotics \cite{Marsano:2012yc} when global Wilson lines are used.

However, the newly gained insights into Abelian symmetries made the direct engineering of the MSSM another valid option \cite{Choi:2013hua,Lin:2014qga,Cvetic:2015txa,Lin:2016vus }. But similar as in the GUT picture, the MSSM gauge group per se is not enough to forbid various dangerous proton decay inducing operators and must be considered incomplete.
One possibility is to extend the symmetries by another gauged $U(1)$ \cite{Lin:2014qga, Lin:2016vus}, however, one is then faced with the issue of how to lift the additional massless photon from the spectrum.
Alternatively, one can add a discrete symmetry, which from the effective field theory perspective is rather minimally invasive and unproblematic, due to the lack of strong anomalies. 
However this is not true anymore when coupled to (quantum) gravity, as here black hole arguments \cite{Krauss:1988zc,Banks:2010zn} suggest an inconsistency of every symmetry that does not have a gauged origin. 
Hence, if we couple gravity to the MSSM, we can only add very specific discrete symmetries that allow a gauging. 
One of the mildest additions to the MSSM that forbids baryon and lepton number violating operators at the renormalizable level which is also known to be (discrete) anomaly free \cite{Ibanez:1991hv} is matter parity.
Therefore, a valid quest is to explore F-theory constructions of such models.

Fortunately, engineering consistent discrete symmetries in F-theory has advanced significantly \cite{Morrison:2014era,Anderson:2014yva,Mayrhofer:2014haa,Mayrhofer:2014laa, Garcia-Etxebarria:2014qua,Cvetic:2015moa} by considering genus-one fibrations. 
In such a setup, the fibration lacks any globally defined section, but instead has only multi-sections, which nevertheless give rise to a well-defined F-theory vacuum.
Indeed, we will demonstrate that in this framework, the MSSM with an additional $\bbZ_2$ symmetry can be engineered straightforwardly with toric methods \cite{Candelas:1996su, Bouchard:2003bu, Braun:2013nqa, Garcia-Etxebarria:2014qua, Mayrhofer:2014haa}.
Moreover, using recent techniques for flux computations \cite{Lin:2015qsa, Lin:2016vus, Bies:2017fam}, we are able to find multiple consistent vacua that exhibit the exact chiral spectrum of the MSSM.
It is worth noting that these techniques are only available because the genus-one fibrations we study have crepant resolutions.
If one limits oneself to only studying F-theory using Weierstrass models, one would need to first develop methods of computing fluxes on singular manifolds, since the Weierstrass models associated with our genus-one fibrations has terminal singularities \cite{Braun:2014oya, Mayrhofer:2014laa, Arras:2016evy, Grassi:2018rva}.

This article is structured as follows: In Section~\ref{sec:two} we review the need for additional discrete symmetries in the MSSM and their field theory constraints from anomaly considerations. 
In Section~\ref{sec:three} we present an F-theory model with matter parity, together with the relevant computational details regarding the geometry and $G_4$-fluxes.
We then demonstrate that these rather formal techniques, when applied to the most simplistic fibrations over the base $B = \bbP^3$, can produce a number of different, consistent configurations that has the exact chiral MSSM spectrum.
Along the way, we highlight the subtle interplay between discrete gauge anomalies in the field theory and intersection number arithmetic of the geometry.
In Section~\ref{sec:four} we construct another fibration in which we will look for other parity assignments, and confront these with discrete anomaly cancellation conditions.
Section~\ref{sec:five} summarizes the results and gives an outlook onto possible future directions.

\section{Prelude: R-parity violation in the MSSM}
\label{sec:two}
In this section we provide a short summary of R-parity violating operators in the MSSM and their tension with experimental bounds. 
These operators have also appeared in earlier F-theory constructions that directly engineer the MSSM \cite{Choi:2013hua,Lin:2014qga,Cvetic:2015txa,Lin:2016vus}.
We review possible $\mathbb{Z}_2$ symmetries that can forbid these operators.

At the renormalizable level, the superpotential of the MSSM can be written as
\begin{align}
\begin{split}
	\mathcal{W}_\text{MSSM} = \, &Y_{i,j}^u Q \overline{u} H_u + Y_{i,j}^d Q \overline{d} H_d + Y_{i,j}^e \overline{e} L H_d + \mu H_u H_d  \\
	& \, + \beta_i L_i H_u + \lambda_{i,j,k} \overline{e} L L + \lambda'_{i,j,k} Q \overline{d} L + \lambda''_{i,j,k} \overline{u}\overline{d}\overline{d} \, .
\end{split}
\end{align}
The couplings in the second row violate baryon and lepton number conservation and are severely constrained by experimental bounds. These usually come from proton or lepton decays mediated by massive superpartners. The strongest bounds are related to products of the above couplings and less strong for the single ones.
Roughly, they are as follows:
\begin{align}
\label{eq:constraints}
\begin{split}
\lambda'' &<  10^{-(2-6)} \left( \frac{ \widetilde{m}  }{100 \text{GeV}} \right)  \, , \qquad  |\lambda \lambda''| < 10^{-(3-19)} \left(\frac{\widetilde{m} }{100 \text{GeV}} \right)^2\, , \\
\lambda' &<  0.01     \left( \frac{ \widetilde{m} }{100 \text{GeV}} \right) \, ,\qquad \qquad |\lambda' \lambda''| < 10^{-27}  \left(\frac{ \widetilde{m}  }{100 \text{GeV}} \right)^2 \, ,\\
\lambda &<  0.01     \left(\frac{ \widetilde{m} }{100 \text{GeV}}\right)  \, , 
\end{split}
\end{align}
where $\widetilde{m}$ are the masses of the decay mediating superpartners generated upon SUSY breakdown (see \cite{Barbier:2004ez,Kao:2009fg} and references therein). 
These couplings can effectively be forbidden by imposing a discrete symmetry on the MSSM which in the simplest case is a $\mathbb{Z}_2$. 
Differing by the discrete charge assignments of MSSM superfields, the phenomenologically most relevant cases are matter, lepton and baryon parity, as summarized in Table~\ref{tab:Z2charges}.
%Note that matter parity admits two different charge assignments related by a $U(1)_Y$ hypercharge rotation. 
\begin{table}[ht]
\begin{center}
	\renewcommand{\arraystretch}{1.3}

	\begin{tabular}{l r}

		\begin{tabular}{|cc|} \hline
			 $G_\text{SM}$ Rep & Matter \\ \hline 
			 $(\mathbf{\mathbf{3},\mathbf{2}})_{\frac16}$& $Q$ \\ \hline
			 $(\mathbf{\overline{\mathbf{3}},\mathbf{1}})_{-\frac23}$& $\overline{u}$\\ \hline
			 $(\mathbf{\overline{\mathbf{3}},\mathbf{1}})_{\frac13}$&$ \overline{d}$ \\ \hline
			 $(\mathbf{\mathbf{1},\mathbf{2}})_{\pm \frac12}$&$ L, H_d,H_u$ \\ \hline
			 $(\mathbf{\mathbf{1},\mathbf{1}})_{1}$ & $e$ \\ \hline
		\end{tabular}

	&

	\begin{tabular}{c|c|c|c|c|c|c|c|}\cline{2-8}
		 & $Q$ & $\overline{u}$ & $\overline{d}$ & $L$ & $\overline{e}$ & $H_d$ & $H_u$ \\ \hline
		\multicolumn{1}{|c|}{$\mathbb{Z}^{M_1}_2$} & + & - & - & + & - & - & - \\ \hline
		\multicolumn{1}{|c|}{  $\mathbb{Z}^{M_2}_2$}   & - & - & - & - & - & + & + \\ \hline
		\multicolumn{1}{|c|}{$\mathbb{Z}^L_2$}   & + & + & + & - & - & + & + \\ \hline
		\multicolumn{1}{|c|}{ $\mathbb{Z}^B_2$ }  & + & - & - & - & + & - & - \\ \hline
	\end{tabular}

\end{tabular}
\caption{\label{tab:Z2charges} Summary of gauge quantum numbers of chiral MSSM superfields. The left table shows 
the gauged quantum numbers, while the table on the right shows the $\mathbb{Z}_2$ charge assignments for matter parities $Z_2^{M_1}$, $Z_2^{M_2}$, lepton parity $\mathbb{Z}_2^L$ and bayron parity $\mathbb{Z}_2^B$.}
\end{center}
\end{table} 
For each charge assignment we summarize the field theoretically forbidden tree level couplings in Table~\ref{tab:YukawaSummary}. 
\begin{table}[ht]
	\begin{center}
		\renewcommand{\arraystretch}{1.2}
		\begin{tabular}{cc|c|c|c|c}
			& Coupling & $\mathbb{Z}_2^{M_1}$ & $\mathbb{Z}_2^{M_2}$& $\mathbb{Z}_2^L$ &$ \mathbb{Z}_2^B$ \\ \hline
			\multirow{3}*{\begin{tabular}{c}Yukawa-\\ Couplings \end{tabular}} & $Q \overline{u} H_u$ & \checkmark & \checkmark &  \checkmark & \checkmark  \\ 
			& $Q \overline{d} H_d$ & \checkmark & \checkmark  &  \checkmark  & \checkmark \\  
			& $\overline{e} L H_d$ & \checkmark & \checkmark  &  \checkmark  & \checkmark \\ \hline 
			$\mu$-term & $H_u H_d$ & \checkmark & \checkmark &  \checkmark & \checkmark \\ \hline
			\multirow{4}*{\begin{tabular}{c} B-\& L- \\ Violation\end{tabular}} & $L H_u$ & X & X & X & \checkmark         \\   
			& $Q \overline{d} L$ & X & X & X & \checkmark\\
			& $\overline{e} L L$ & X & X & X & \checkmark \\
			& $\overline{u}\overline{d}\overline{d}$ & X & X& \checkmark & X \\ \hline
		 \end{tabular}
	 \caption{\label{tab:YukawaSummary} Summary of allowed and forbidden tree level couplings under different $\mathbb{Z}_2$ symmetry charge assignments. While the first four terms are required to be present the later four should better be forbidden.}
	\end{center}
\end{table}
Both matter parities can forbid all unwanted baryon and lepton number violating couplings while lepton and baryon parity can only forbid their respective ones.

Two comments concerning the discrete charge assignments are in order. 
First we note that the two matter parities are field theoretically equivalent by mixing the $U(1)_Y$ charge with the $\mathbb{Z}_2$. Explicitly, we have
\begin{align}\label{eq:rotation_matter_parity}
{\bbZ}^{M_1}_2 =   {\bbZ}^{M_2}_2 + 6 \, U(1)_Y \, \, \text{ mod } 2 \, .
\end{align}
As all $SU(3)$ triplets have hypercharges that are multiples of 1/3, the above redefinition is trivial for them.
Meanwhile, $SU(2)$ doublets as well as the bifundamental states have hypercharge 1/2 or 1/6, which leads to a sign flip upon performing the rotation \eqref{eq:rotation_matter_parity}.
%In Sections~\ref{sec:three} and \ref{sec:four} we construct both charge assignments and comment on their possible equivalence.
Second we note that matter parity is clearly superior to the other $\mathbb{Z}_2$ charge assignments when it comes to the problematic tree level couplings. However, in SUSY breaking schemes where the sfermion masses are large, the other charge assignments might still be phenomenologically interesting. 
In such cases certain individual couplings might still be within their mild experimental bounds and thus acceptable.
Only the products in \eqref{eq:constraints} with the $\lambda''$ coupling are strongly constrained and those are still forbidden for both parities.

\subsection*{Discrete Anomalies}

From a phenomenological perspective, a UV completion of the MSSM with gravity might appear to be a purely ideological endeavor.
However, if we include an effective global discrete symmetry, the subtle interplay with gravity could affect even low energy physics.
Indeed, it has been argued that global symmetries can be broken by quantum gravity effects \cite{Krauss:1988zc,Kallosh:1995hi,Banks:2010zn }, thus re-introducing the proton decay inducing operators.
The only exception are discrete gauge symmetries, i.e., remnants of broken gauge symmetries.
This on the other hand puts constraints on the spectrum and charges of $\mathbb{Z}_n$ symmetries as they must be embeddable into an anomaly free Abelian gauge group.
In the case $n=2$, the non-trivial conditions for an anomaly free discrete symmetry that are inherited from the unbroken phase are \cite{Ibanez:1991hv}:
\begin{align}\label{eq:discrete_anomaly_cancellation_general}
	\begin{split}
		 \bbZ_2 - G^2 :   \, \sum_{{\bf R}_G} Q_{\bbZ_2}({\bf R}) \, C^{(2)}_{G} ({\bf R})= m\, , \quad  m \in \mathbb{Z} \, ,
	\end{split}
\end{align}
where $G$ is the non-Abelian gauge group.\footnote{There are also mixed Abelian discrete symmetries, that are less conclusive due to ambiguities in the charge normalization \cite{Ibanez:1991hv}.}
Here $C^{(2)}_{G} ({\bf R})$ is the quadratic Casimir invariant of the representation $\mathbf{R}$, normalized such that is $1/2$ for the fundamental representation.
As found in \cite{Ibanez:1991hv, Ibanez:1991pr} and can be easily checked with the spectrum in Table \ref{tab:Z2charges}, out of the above $\mathbb{Z}_2$ symmetries, only the two matter parities are anomaly free with just the MSSM spectrum, and hence have a corresponding gauge origin in the UV without chiral exotics. 
We will see in the next section that in global F-theory models, where gravity is automatically coupled to the gauge sector, these subtle field theoretic constraints reappear as geometric consistency conditions.

\section{F-theory construction of 4d MSSM vacua with matter parity}
\label{sec:three} 
In this section, we will present the details of an F-theory compactification which realizes a three-family MSSM vacuum with an additional $\bbZ_2$ symmetry that is identified with the matter parity $\bbZ_2^{M_2}$.
The necessary ingredients, which we will now discuss in the same order, are
\begin{enumerate}
	\item the generic fiber structure that realizes the Abelian part of the gauge symmetry,
	\item codimension one singularities (with resolution) corresponding to the non-Abelian gauge algebra,
	\item matter representations associated with codimension two fiber components and $G_4$-fluxes,
	\item specification of the base and fibration and consistent three-family configurations.
\end{enumerate}
By keeping the base generic for the first three points, we will have the capabilities to analyze a large number of concrete models for the last point.

\subsection{Toric hypersurface with two bisection classes}
 
With our phenomenological motivations, we seek to realize an F-theory model whose Abelian gauge sector is $U(1) \times \bbZ_2$.
As studied in \cite{Klevers:2014bqa}, a straightforward fiber type that does the job is described in terms of one of the 16 reflexive 2D polygons.
Labelled as $F_2$ in \cite{Klevers:2014bqa}, the generic fiber $\mathfrak{f}$ of this geometry is a bi-quadric curve, given as the vanishing of the polynomial
\begin{align}\label{eq:toric_P1P1_hypersurface}
\begin{split}
	p_{F_2} & = (b_1\,y^2 + b_2\,s\,y + b_3\,s^2) \, x^2 + (b_5\,y^2 + b_6\,s\,y + b_7\,s^2)\,x\,t + (b_8\,y^2 + b_9\,s\,y + b_{10}\,s^2)\,t^2  
\end{split}
\end{align}
inside the surface $\bbP^1 \times \bbP^1$ with homogeneous coordinates $([x:t], [y:s])$.
By promoting the coefficients $b_i$ to functions over a complex threefold base $B$, we obtain a genus-one fibered fourfold $Y \subset \cA$.
Here, $\cA$ is the ambient space obtained by fibering $\bbP^1 \times \bbP^1$ over the same base $B$.
The full configuration is summarized via the commutative diagram:
\begin{equation}
	\begin{tikzcd}[row sep = normal, column sep= normal  ]
		\mathfrak{f} \arrow[hook, r] \arrow[hook, dr] & \bbP^1 \times \bbP^1 \arrow[hook, r] & \cA \arrow[d] \\
		& Y \arrow{r}{}[swap]{\pi} \arrow[ru, hook] & B
	\end{tikzcd} \, .
\end{equation}
The ambient space fibration is specified by two line bundles with divisor classes $S_7$ and $S_9$ over the base.
They determine the relative ``twisting'' of the fiber coordinates over the base via the linear equivalence relations
\begin{align}\label{eq:bisection_classes_F2_noTop}
	[x] = [t] - \KB + S_9 \, , \quad [y] = [s] - \KB + S_7 \, ,
\end{align}
where $\KB$ is the anti-canonical class of the base.
For $Y$ to be Calabi--Yau, the coefficients $b_i$ of the polynomial \eqref{eq:toric_P1P1_hypersurface} have to be sections with the following divisor classes:
\begin{align}\label{eq:coefficients_F2_classes_noTop}
	\begin{array}{lll}
		[b_1] = 3\,\KB - S_7 - S_9 \, , &  [b_2] = 2\,\KB - S_9 \, , & [b_3] = \KB + S_7 - S_9 \, , \\ [1ex] %
		[b_5]= 2\,\KB - S_7 \, , & [b_6] = \KB \, , & [b_7] = S_7  \, , \\[1ex]
	 	[b_8] = \KB + S_9 - S_7 \, , & [b_9] = S_9 \, , & [b_{10}] = S_7 + S_9 - \KB \, .
	\end{array}
\end{align}

The genus-one fibration $Y$ has no rational section, but two independent bisection classes \eqref{eq:bisection_classes_F2_noTop}, corresponding to the two hyperplanes of the fiber ambient space $\bbP^1 \times \bbP^1$ (modulo vertical divisors).
They have been shown to give rise to a $U(1) \times \bbZ_2$ symmetry in F-theory \cite{Klevers:2014bqa,Grimm:2015wda}.
A multisection of a genus-one fibration is just as good as a section of an elliptic fibration when it comes to identifying the Kaluza--Klein (KK) $U(1)$ in the dual M-theory compactification \cite{Braun:2014oya}.
However, in the absence of a section, new subtleties arise in the additional possibilities that an $n$-section can intersect codimension two fiber components.
As has been extensively studied in \cite{Anderson:2014yva, Garcia-Etxebarria:2014qua, Mayrhofer:2014haa, Mayrhofer:2014laa, Cvetic:2015moa}, one can understand these intersection numbers as the mod $n$ charge of the matter states under a discrete $\bbZ_n$ symmetry, which in the M-theory phase is mixed with the KK-$U(1)$ to give rise to the massless Abelian vector field dual to the $n$-section class.

For the model \eqref{eq:toric_P1P1_hypersurface}, we pick the divisor class 
\begin{align}\label{eq:Z2_gen_noTop}
	D_{\bbZ_2} := [x]
\end{align} to be the one dual to the massless vector of the KK/$\bbZ_2$ combination.
Then the other bisection class, $[y] = [s] \text{ mod } D_B$, gives rise to another massless vector that uplifts to a genuine massless $U(1)$ gauge field in F-theory.
To be precise, this $U(1)$ is dual to a divisor class that is ``orthogonal'' to $D_{\bbZ_2}$, which is
\begin{align}\label{eq:U1_gen_noTop}
	D_{U(1)} = [y]- [x] + \frac{1}{2} (\KB + S_7 - S_9) \, .
\end{align}
This modified divisor can be understood as a generalized Shioda map for genus-one fibrations with more than one independent multisection class \cite{Grimm:2015wda}.

As extensively studied in \cite{Klevers:2014bqa}, this bisection geometry can be obtained through a complex structure deformation of an elliptic fibration with Mordell--Weil rank two, which in F-theory gives rise to a $U(1)^2$ gauge group.
The associated conifold transition corresponds in field theory to the Higgsing of one of these $U(1)$ factors to a $\bbZ_2$ by giving vev to a singlet of charge $(0,2)$.
Therefore, we see explicitly that the $\bbZ_2$ symmetry we construct via F-theory has a gauged origin, and hence should have a consistent quantum gravity embedding.

\subsection{Non-Abelian symmetries with matter parity via tops}

We include non-Abelian $SU(3) \times SU(2)$ gauge symmetries using the methods of tops \cite{Candelas:1996su,Bouchard:2003bu}.
There are four possibilities for each of the $SU(3)$ and $SU(2)$ tops, cf.~appendix \ref{app:B}.
In the following, we will focus on a combination of $SU(3)$ top 3 and $SU(2)$ top 1.
The toric description modifies the ambient space $\cA$ to include additional toric divisors $\{f_i\}_{i=0,1,2}$ and $\{e_j\}_{j=0,1}$, which themselves are fibered over codimension one loci $\{w_3\}$ resp.~$\{w_2\}$ inside the base $B$.
Their restriction, or intersection, with the likewise modified hypersurface $\hat{Y} \equiv Y_{31}$ are now the exceptional, or ``Cartan'' divisors that resolve the $SU(3)$ resp.~$SU(2)$ singularities over respective codimension one loci on $B$.
The $\bbP^1$ fibers of these divisors intersect in the affine Dynkin diagrams of the corresponding Lie algebra.

Explicitly, the modified hypersurface equation is given by the vanishing of a polynomial $p_{31}$, which is a specialization of the polynomial \eqref{eq:toric_P1P1_hypersurface} with coefficients
\begin{align}\label{eq:coefficients_top3x1}
\begin{split}
	&b_ 1 = d_ 1\, e_ 0\, f_ 1,  \quad b_ 2 = d_ 2\, e_ 0\, f_ 0\, f_ 1, \quad b_ 3 = d_ 3\, e_ 0\, f_ 0^2\, f_ 1, \quad b_ 5 = d_ 5\, f_ 1\, f_ 2, \\
	& b_ 6 = d_ 6, \quad b_ 7 = d_ 7\, f_ 0, \quad b_ 8 = d_ 8\, e_ 1\, f_ 1\, f_ 2^2,  \quad b_ 9 = d_ 9\, e_ 1\, f_ 2, \quad b_ {10} = d_ {10}\, e_ 1\, f_ 0\, f_ 2 \, .
\end{split}
\end{align}
The functions $d_i$ are again sections of line bundles over the base, whose divisor classes are related to those without the top \eqref{eq:coefficients_F2_classes_noTop} via
\begin{align}\label{eq:coefficients_F2_classes_top3x1}
	\begin{array}{lll}
		[d_1] =[b_1] - W_2 \, , &  [d_2] = [b_2] - W_2 - W_3 \, , & [d_3] = [b_3] - W_2 - 2\,W_3 \, , \\ [1ex] %
		[d_5]= [b_5] \, , & [d_6] = [b_6] \, , & [d_7] = [b_7] - W_3  \, , \\[1ex]
	 	[d_8] = [b_8] \, , & [d_9] = [b_9] \, , & [d_{10}] = [b_{10}] - W_3 \, ,
	\end{array}
\end{align}
where we have denoted the classes of $\{w_{2/3}\}$ by $W_{2/3}$.
Furthermore, we shall denote the classes of the exceptional divisors by $F_i$ resp.~$E_j$.
Though these are strictly speaking classes on the ambient space $\cA$, we will abusively use the same notation for their restrictions to $Y_{31}$.

Through the toric construction, we can straightforwardly determine the  linear equivalence relations (LIN) between the divisors and the Stanley--Reisner ideal (SRI), that is, the set of divisors whose intersection product is trivial in the Chow ring.
For $Y_{31}$, these are
\begin{equation}
	\begin{split}
		\text{LIN} = \{ &  [x] = [t] + E_1 + F_2 -\KB  + S_9,   \, W_2 = E_0 + E_1  \, , \\
      	& [s] = [y] + F_1 + F_2 + \KB  - S_7,   \, W_3 = f_0 + f_1 + f_2   \} \, ,\\
      	\text{SRI} =\{ &xt,xe_1,xf_2,ys,yf_0,te_0,e_0f_2,tf_1,sf_1,sf_2\} \, .
     \end{split}
\end{equation}
In the presence of codimension one reducible fibers, the divisors dual to the KK/$\bbZ_2$ and the $U(1)$ vector field needs to be refined, in order for these to be ``orthogonal'' to each other and to the Cartan $U(1)$s of the non-Abelian symmetries.
Geometrically, this is necessary because the bisections will intersect the fibers of the exceptional divisors non-trivially.
For example, while the bisection $[x]$ intersects only the affine node (the fiber component of $E_0$) of the $SU(2)$, it intersects the $SU(3)$ divisors non-trivially, namely each of the fibers of $F_0$ and $F_1$ once.
Physically, it would mean that the W-bosons of $SU(3)$ were charged non-trivially under the $\bbZ_2$, which is of course unacceptable.
However, much like in the case of $U(1)$s, we can add a linear combination of the Cartan divisors to the (bi-)section to correct the intersection numbers \cite{Garcia-Etxebarria:2014qua, Mayrhofer:2014haa, Lin:2015qsa, Buchmuller:2017wpe}.
For the case at hand, it can be checked that the correct $\bbZ_2$ is given by the divisor
\begin{align}\label{eq:Z2_gen_top3x1}
	D_{\bbZ_2} = [x] + \frac{1}{3}(2\,F_1 + F_2) \, .
\end{align}
Similarly, the modified $U(1)$ generator that is orthogonal to the Cartans as well a the $\bbZ_2$ is
\begin{align}\label{eq:U1generator_top3x1}
	D_{U(1)} = [x] - [y] - \frac{1}{2}\,E_1 - \left( \frac{1}{3}\,F_1 + \frac{2}{3}\,F_2 \right) - \frac{1}{3}\,W_3 + \frac{1}{2}\,\KB + \frac{1}{2}\,S_7 - \frac{1}{2}\,S_9 \, .
\end{align}
Note that we have also flipped the sign of the bisections compared to \eqref{eq:U1_gen_noTop}, so that it matches the hypercharge $U(1)_Y$ of the MSSM.

\subsubsection{The global gauge group structure} 

Let us briefly discuss the global group structure of this model.
Though our fourfold $Y_{31}$ is not elliptically fibered, we can apply the same intersection number argument employed in \cite{Cvetic:2017epq} to determine the constraints on the charges and non-Abelian representations of matter states arising from M2-branes wrapping fibral curves $\Gamma$.
Essentially, one employs the fact that multisections as integer divisor classes have integer intersection number with any fiber component.
This in turns means that the Abelian charges---given by intersection numbers of $\Gamma$ with the divisors \eqref{eq:Z2_gen_top3x1} and \eqref{eq:U1generator_top3x1}---differ by an integer from the specific fractional linear combinations of the non-Abelian weights---given by the fraction linear combination of exceptional divisors in \eqref{eq:Z2_gen_top3x1} and \eqref{eq:U1generator_top3x1}.
For the $U(1)$, the fractional contributions from the exceptional divisors of both $SU(3)$ and $SU(2)$ are exactly those compatible with the $\bbZ_6$ quotient of the continuous part of the gauge algebra \cite{Cvetic:2017epq}, namely charge $1/2 \text{ mod }\bbZ$ for doublets, $2/3 \text{ mod }\bbZ$ for triplets, and $1/6 \text{ mod } \bbZ$ for bifundamentals.

For the discrete symmetry, note that triplets will generically have charges quantized in $1/3$ with respect to the divisor in \eqref{eq:Z2_gen_top3x1}.
Because the generic fiber still has intersection 2 with $D_{\bbZ_2}$, we can still only interpret the charge under $D_{\bbZ_2}$ modulo 2.
Thus, naively, we would expect that the discrete symmetry is enhanced to a $\bbZ_6$ by the presence of the non-Abelian symmetries.
But not all charges of the $\bbZ_6$ can appear!
First, it is clear that $SU(2)$ matter will only be charged under a $\bbZ_2$ subgroup, because their intersection numbers with $D_{\bbZ_2}$ are integer.
Furthermore, by the analogous argument as in \cite{Cvetic:2017epq}, one can construct an order three central element of $SU(3) \times \bbZ_6$ which acts trivially on any matter states.
Essentially, it follows because $(D_{\bbZ_2} - 1/3(2F_1 + F_2) ) \cdot \Gamma = [x] \cdot \Gamma \in \bbZ$.
Since $\bbZ_6 = \bbZ_2 \times \bbZ_3$ has a unique order three subgroup, we conclude that the only non-trivially acting part is the $\bbZ_2$.
To infer the charges under it, we can simply multiply all intersection numbers with $D_{\bbZ_2}$ with three and then take the result modulo 2.
In order to differentiate it more easily from the $U(1)$ charges, we will denote even/odd charges by $+/-$.
To summarize, the global gauge group of the F-theory model on $Y_{31}$ is
\begin{align}
	\frac{SU(3) \times SU(2) \times U(1)}{\bbZ_6} \times \bbZ_2 \, .
\end{align}

Note that it was already anticipated before in \cite{Anderson:2014yva} that, by engineering a non-Abelian symmetry algebra $\mathfrak{g}$ inside an $n$-section fibration, the discrete symmetry can be potentially enhanced to $\bbZ_{n \times r}$, where $r$ is the order of the center $Z(\mathfrak{g})$.
In general, $\bbZ_{n \times r} \neq \bbZ_n \times \bbZ_r$ (namely, whenever $n$ and $r$ are \textit{not} coprime), and the enhancement can be physical.
However, due to the mechanism that leads to such an enhancement---namely, the divisor associate with the discrete symmetry is shifted by the Cartan divisors---the resulting global gauge group necessarily has to be non-trivial.
For example, there is an $SU(2)$ top constructed over the $F_2$ polygon that has an enhancement, such that the gauge group is $[SU(2) \times \bbZ_4]/\bbZ_2$ (we have omitted the $U(1)$, which itself has a non-trivial gauge group structure associated with the $SU(2)$, see Table \ref{tab:su2top3}).
We will leave a detailed derivation and classification along the lines of \cite{Cvetic:2017epq} for future work.

\subsection{Matter surfaces, fluxes and the chiral spectrum}

To specify the chiral spectrum of the F-theory compactification, we need two geometric ingredients, namely the surfaces on which the matter states are localized, and the description of the $G_4$-flux in terms of their dual four cycle classes.
Based on the techniques first developed in \cite{Krause:2011xj, Krause:2012yh} and further advanced in \cite{Cvetic:2013uta, Cvetic:2015txa, Lin:2015qsa, Lin:2016vus, Bies:2017fam}, we perform a completely base independent analysis of fluxes and chiralities, which then can be straightforwardly applied to specific fibrations.

\subsubsection{Matter surfaces and their homology classes}

Through the mapping to its Jacobian \cite{Klevers:2014bqa}, we can straightforwardly determine the codimension two loci where the fiber singularities of $Y_{31}$ enhance.
These are of them form $\{w_i\} \cap \{g_{\bf R} \}$ for some polynomials $g_{\bf R}$:
\begin{equation}\label{eq:matter_curves_equations_top3}
    \begin{split}
        &g_{{\bf 2}_1}=d_{10}d_6^2d_8 - d_{10}d_5d_6d_9 - d_6d_7d_8d_9 + d_5d_7d_9^2 + w_3 \, (d_{10}^2d_5^2  - 2d_{10}d_5 + d_7^2d_8^2 ) \, ,\\ 
        &g_{{\bf 2}_2}=d_1d_3d_6^2 - d_1d_2d_6d_7 + d_1^2d_7^2  + w_3\,( d_2^2d_5d_7 - d_2d_3d_5d_6  - 2d_1d_3d_5d_7 + d_3^2d_5^2 w_3) \, ,\\
        &g_{{\bf 3}_1}=d_1 \, ,\\
        &g_{{\bf 3}_2}=d_{10}d_6 - d_7d_9 \, ,\\
        &g_{{\bf 3}_3}=d_3d_6^2 - d_2d_6d_7 + d_1d_7^2 \, ,\\
        &g_{{\bf 3}_4}=d_6^2d_8 - d_5d_6d_9 + d_1d_9^2 w_2 \, .
    \end{split}
\end{equation}
Furthermore, there are two charged singlets with $U(1)$ charge 1, but differ in their $\bbZ_2$ charge, which are localized over curves given by complicated ideals $I_{+/-}$.

There is also an uncharged singlet with negative $\bbZ_2$ parity \cite{Klevers:2014bqa}.
These matter states have the same quantum numbers as right-handed neutrinos.
Correspondingly, they interact with Higgs and lepton doublets via perturbatively realized Yukawa couplings in the F-theory geometry.
The presence of such massless states in the effective field theory not only depends on the flux, but also on the complex structure moduli \cite{Bies:2014sra, Bies:2017fam}.
However, because it is a real representation, there cannot be any net chirality associated with these matter states.
Geometrically, this is reflected in the fact that the components of the I$_2$ fiber associated with this matter are exchanged via monodromy.
Consistently, the transversality conditions \eqref{eq:transversality_conditions} imply that the intersection product between the flux and this singlet's matter surface is 0.
Hence, we will disregard this representation for the rest of this paper, since we are mainly interested in the chiral spectrum.

Over the codimension two loci $C_{\bf R} = \{w_i\} \cap \{g_{\bf R}\}$, the reducible fibers contain localized $\bbP^1$ components, giving rise to matter states in the representation ${\bf R}$ in F-theory.
By fibering one such $\bbP^1$ over the curve on the base, one obtains a four-cycle $\gamma_{\bf R}$, a so-called matter surface associated with a weight $\bf w$ of a representation ${\bf R}$ (or its conjugate).
To determine the homology classes of these matter surfaces, we use prime ideals techniques and algorithms detailed in \cite{Lin:2016vus}, utilizing the computer algebra program \texttt{Singular} \cite{DGPS}.
The specific states, whose matter surfaces we determine this way, are listed in Table \ref{tab:matter_surfaces_charges_top3}.
Their corresponding matter surface classes are collected in the appendix, cf.~Table \ref{tab:top3x1_matter_homology_class}.

\begin{table}[ht]
\begin{align*}
	\begin{array}{>{\displaystyle} c | >{\displaystyle} c  | >{\displaystyle} c | >{\displaystyle} c | >{\displaystyle} c | >{\displaystyle} c}
		\text{rep} & U(1) \times \mathbb{Z}_2 & \text{Cartan charges} & \text{parent exceptional} & \text{base locus} & \text{$\mathbb{Z}_2^{M_2}$ Rep.}\\ \hline \hline \rule{0pt}{4ex}
		{\bf 2}_1 & \left( \frac{1}{2}, - \right) & (0,0 \, | \, 1) & E_0 & \{w_2\} \cap \{g_{{\bf 2}_1}\} & L \\ \rule{0pt}{4ex}
		{\bf 2}_2 & \left( \frac{1}{2}, + \right) & (0,0 \, | \, -1) & E_1 & \{w_2\} \cap \{g_{{\bf 2}_2}\} & \text{Higgs} \\ \rule{0pt}{4ex}
		\overline{\bf 3}_1 & \left( -\frac{2}{3}, - \right) & (0,1 \, | \, 0) & F_0 & \{w_3\} \cap \{g_{{\bf 3}_1}\} & \bar{d} \\ \rule{0pt}{4ex}
		{\bf 3}_2 & \left( \frac{2}{3}, + \right) & (-1,1 \, | \, 0) & F_1 & \{w_2\} \cap \{g_{{\bf 3}_2}\} & \text{exotic} \\ \rule{0pt}{4ex}
		\overline{\bf 3}_3 & \left( \frac{1}{3}, - \right) & (1,-1 \, | \, 0) & F_2 & \{w_3\} \cap \{g_{{\bf 3}_3}\} & \bar{u} \\ \rule{0pt}{4ex}
		\overline{\bf 3}_4 & \left( \frac{1}{3}, + \right) & (0,1 \, | \, 0) & F_0 & \{w_3\} \cap \{g_{{\bf 3}_4}\} & \text{exotic} \\ \rule{0pt}{4ex}
		{\bf (\overline{3},2) } & \left( - \frac{1}{6}, - \right) & (0,1 \, | \, -1) & E_1, \, F_0& \{w_2\} \cap \{w_3\} & Q \\ \rule{0pt}{4ex}
		{\bf 1}_{1} & \left(1, -\right) & (0, 0 \, | \, 0) & \text{n.a.} & V(I_{(1,-)}) & E \\ \rule{0pt}{4ex}
		\overline{\bf 1}_{2} & \left(-1, + \right) & (0 , 0 \, | \, 0) & \text{n.a.} & V(I_{(1,+)} ) & \text{exotic} \\ \rule{0pt}{4ex}
		{\bf 1}_3 & \left( 0 , - \right) & (0 ,0 \, | \, 0) & \text{n.a.} & V(I_{(0,-)}) & \text{exotic} \,  (\nu_R)
	\end{array}
\end{align*}
\caption{States and charges associated with the matter surfaces. The polynomials $g_i$ defining the matter curves in the base are in equation \eqref{eq:matter_curves_equations_top3}. We have included the identification with the MSSM spectrum, where the $\bbZ_2$ is identified with matter parity.}\label{tab:matter_surfaces_charges_top3}
\end{table} 
Note that the only bifundamental matter states we have in this model have odd $\bbZ_2$ charge.
Therefore, this toric model $Y_{31}$ only allows for an identification of the geometrically realized $\bbZ_2$ with the second matter parity $\bbZ_2^{M_2}$ listed in Table \ref{tab:Z2charges}.
We will see in the next section other geometries whose corresponding F-theory model may realize the other parities. 
\subsubsection{Vertical fluxes from matter surfaces} 
We now turn to the computation of $G_4$-fluxes.
In practice, these are expressed through their Poincar\'{e}-dual four cycle classes (also denoted by $G_4$), such that the integral giving the chiral index can also be computed via intersection product:
\begin{align}\label{eq:basic_formula_chirality}
	\chi ({\bf R}) = \int_{\gamma_{\bf R}} G_4 = G_4 \cdot [\gamma_{\bf R}]
\end{align}
Not all four cycles give rise to consistent fluxes.
As is well known by now, the fluxes have to satisfy the so-called transversality conditions \cite{Dasgupta:1999ss} in order to uplift from M- to F-theory.
These conditions have been generalized in \cite{Lin:2015qsa} to genus-one fibrations:
\begin{align}\label{eq:transversality_conditions}
	G_4 \cdot D_B^{(1)} \cdot D_B^{(2)} =0 \, , \quad G_4 \cdot D_B \cdot [x] = 0
\end{align}
for some vertical divisors $D_B^{(i)}$.
In addition, the flux must not break the non-Abelian symmetries, which requires
\begin{align}\label{eq:gauge_symmetry_condition}
	G_4 \cdot \text{Ex} \cdot D_B = 0 \, ,
\end{align}
where $\text{Ex} \in \{ E_1, F_1, F_2\}$ are the Cartan divisors.

For a fibration over a base threefold $B$, we can use the quotient ring description to determine a basis of vertical fluxes \cite{Lin:2016vus}.
For generic choices of fibration and base, i.e., such that no further singularity enhancements are induced whose resolution would introduce further divisors, the space of vertical fluxes is spanned by $U(1)$-fluxes of the form $D_{U(1)} \cdot F$, where $F \in \pi^* (H^{1,1}(B))$, and five non-$U(1)$-fluxes.
In this paper, we follow the method of \cite{Bies:2017fam} and express the non-$U(1)$-fluxes through so-called matter surfaces fluxes.
At this point, there is no technical advantage for this procedure, and we could also use the flux basis provided by the algorithm of \cite{Lin:2016vus} to compute the chiralities.
However, hoping that in future works we will have the computational power to also determine the vector-like spectrum with the methods of \cite{Bies:2017fam}, we will collect the necessary input in the Appendix \ref{app:C}.
For now, we content ourselves with a basis for vertical fluxes in terms of matter surfaces.

As the name suggests, the matter surface fluxes are constructed using the matter surfaces $[S_{\bf R}]$.
By construction, these surfaces are orthogonal to any curve in the base:
\begin{align}\label{eq:transversality_1_for_matter_surface}
	[S_{\bf R}] \cdot D_B^{(1)} \cdot D_B^{(2)} = 0 \, ,
\end{align}
thus automatically satisfying the first of the transversality conditions \eqref{eq:transversality_conditions}.
To satisfy the other as well as the gauge symmetry condition \eqref{eq:gauge_symmetry_condition}, we can add correction terms of the form $\text{Ex}_i \cdot D_B + D_B^{(a)} \cdot D_B^{(b)}$ to $[S_{\bf R}]$.
Note that these correction terms will not spoil the condition \eqref{eq:transversality_1_for_matter_surface}.
Denoting such corrected matter surfaces by $A({\bf R})$, we can choose a basis of five of them such that together with the $U(1)$-fluxes, they span the full space of vertical fluxes.
Here, we will use the fluxes associated with ${\bf 2}_2, {\bf 3}_2, \overline{\bf 3}_4, {\bf (\overline{3},2)}, \overline{\bf 1}_{2}$, which are:
\begin{align}\label{eq:matter_surface_fluxes_basis}
	\begin{split}
		A({\bf 2}_2) & = [S_{{\bf 2}_2}] + E_1 \cdot (W_2 + W_3 - 3\,\KB + S_9) \, , \\
		A({\bf 3}_2) & = [S_{{\bf 3}_2}] + \frac{1}{3} \, (-F_1 + F_2) \cdot (S_7 + S_9 - W_3) - \frac{1}{3} \, W_3 \cdot [g_{{\bf 3}_2}] \, , \\
		A({\bf 3}_4) & = [S_{\overline{\bf 3}_4}] + \frac{1}{3} \, (F_1 + 2\,F_2) \cdot (3\,\KB - S_7 + S_9) - \frac{2}{3} \, W_3 \cdot [g_{{\bf 3}_4}] \, , \\
		A({\bf (3,2)}) & = [S_{\bf (\overline{3},2)}] + \frac{1}{3}\,(F_1 + 2\,F_2) \cdot W_2 - \frac{1}{2}\,E_1\cdot W_3 - \frac{1}{6}\,W_2 \cdot W_3 \, , \\
		A({\bf \overline{1}}_2) & = [S_{\overline{\bf 1}_{2}}] \, .
	\end{split}
\end{align}
Including the $U(1)$-flux, we parametrize the most generic vertical $G_4$ in this model as
\begin{equation}\label{eq:generic_flux}
    G_4=a_1A(\two_2)+a_2A(\three_2)+a_3A(\three_4)+a_4A(({\three},\two))+a_5A(\overline{\one}_{2})+D_{U(1)}\wedge F \, .
\end{equation}
The chiral indices of matter representations can be straightforwardly computed in the quotient ring description of the vertical cohomology ring.
Instead, one can also use the more geometric picture laid out in \cite{Bies:2014sra, Bies:2017fam} and relate the chiralities to the homology classes of the Yukawa points.
Their rather uninspiring explicit expressions are presented in the appendix, Formula \eqref{eq:chiralities_top1}.
As a consistency check, it is straightforward to verify that all 4D continuous gauge anomalies induced by the chiral spectrum are indeed canceled.
It would be interesting to reproduce this result also in the weakly coupled type IIB limit of this model, along the lines of \cite{MayorgaPena:2017eda}.

\subsection{Concrete three family models}

We are now in a position to scan for configurations that admit a three family flux solution.
Recall that because the bifundamental states in this geometry have odd $\bbZ_2$ charge, the only phenomenological parity extension that is compatible is the matter parity $\bbZ_2^{M_2}$.
Then, consistency with the observed spectrum (cf.~table \ref{tab:Z2charges}) requires to have the following chiral indices of the matter representations:
\begin{align}\label{eq:chirality_condition_top3x1}
\renewcommand{\arraystretch}{1.6}
\begin{array}{|c|c|c|c|c|c|c|c|c|c|}\hline
\textbf{R} & \two_1 & \two_2 & \overline{\three}_1 & \three_2 & \overline{\three}_3 & \overline{\three}_4  & (\overline{\three},\two) & \one_{(1,+)} & \one_{(1,-)} \\ \hline
\chi & -3 & 0 & 3 & 0 & 3 & 0 & -3 & 0 & 3 \\ \hline
\end{array}
\end{align}
In addition to these chiral indices, we have to ensure the vanishing of the flux-induced D-term of the $U(1)$,
\begin{align}
	\xi \sim G_4 \wedge D_{U(1)} \wedge J_B \, ,
\end{align}
where $J_B$ is the K\"ahler form of the base.
Note that this expression can be easily computed when we express $G_4$ in terms of the matter surface fluxes \eqref{eq:matter_surface_fluxes_basis}, because the $U(1)$ generator $D_{U(1)}$ is orthogonal to all the correction terms.
Hence, the D-term is just a linear combination of the matter curves times the base's K\"ahler form, where the coefficients are the $U(1)$ charges.
For the explicit flux parametrization \eqref{eq:generic_flux}, this yields
\begin{align}\label{eq:D-term_U1}
	\xi \sim J_B \cdot \left( \frac{a_1}{2} \, C_{{\bf 2}_2} + \frac{2\,a_2}{3} \, C_{{\bf 3}_2} + \frac{a_3}{3} \, C_{{\bf 3}_2} - \frac{a_4}{6} \, W_2\,W_3 - a_5\,C_{{\bf 1}_2} + \underbrace{\left( \frac12 \, W_2 + \frac23\,W_3 - 2\, \KB \right)}_{ = \pi_*(D_{U(1)} \cdot D_{U(1)})} \, F \right)\, .
\end{align}

To find explicit models with this chiral spectrum, we need to specify the base $B$, the fibration structure in terms of the classes $S_{7/9}$, the choices for the non-Abelian divisors $W_{2/3}$, and the explicit flux which induces the correct chiralities.
To make our lives as easy as possible, we will restrict ourselves to the simplest possible base, $B = \bbP^3$. As we shall see, this choice admits multiple solution and is hence by no means too restrictive.

\subsubsection*{Realistic chiral models over the base $\mathbb{P}^3$}

For this simple choice of the base, the only independent divisor class is the hyperplane class $H$.
Parametrizing the divisors in terms of $H$,
\begin{equation}
\overline{K}_B = 4H \; , \quad  W_2 = n_2H\; , \quad W_3 = n_3H \; , \quad S_7 = s_7H \; , \quad S_9 = s_9H .
\end{equation} 
the integers $n_i, s_j$ have to be such that the classes \eqref{eq:coefficients_F2_classes_top3x1} of the coefficients $d_i$ as well as $W_{2,3}$ are non-negative.
On this specific base, the generic vertical flux \eqref{eq:generic_flux} is parametrized by the five coefficients $a_i$ of the matter surfaces fluxes and the $U(1)$-flux $D_{U(1)} \wedge (\lambda\,H)$.
We collect these numbers in the flux vector $\mathcal{F}=(a_1,a_2,a_3,a_4,a_5,\lambda)$.
We can now scan over all possible fibrations over $B= \bbP^3$ for flux solutions that generate the spectrum \eqref{eq:chirality_condition_top3x1} as well as a vanishing D-term \eqref{eq:D-term_U1} (for $B= \bbP^3$, the K\"ahler form is simply $J_B \sim H$).
In Table~\ref{tab:FluxSummary1}, we list the five configurations satisfying these requirements.
\begin{table}[ht]
 \begin{center}
 \renewcommand{\arraystretch}{1.5}
 \begin{tabular}{|c|c|r c ccccc|c |        }\cline{2-10}
\multicolumn{1}{c|}{} & $ (n_2, n_3, s_7, s_9)$ &flux $\mathcal{F}=$&$(a_1,$&$ a_2,$&$ a_3,$&$ a_4,$&$ a_4,$&$ \lambda)$ & $n_{D3}$ \\ \hline
 \multirow{5}*{$\mathbb{Z}_2^{M_2}$} 
 & $(1,3,7,3)$ & &$(0,$&$ \frac74,$&$-\frac34,$&$-2,$&$\frac34,$&$-\frac32)$ & 33 \\ \cline{2-10}
 &$ (3,1,5,1)$ & &$(0,$&$-\frac{5}{12} ,$&$ -\frac{1}{12} ,$&$-\frac23 ,$&$  \frac{1}{12},$&$-\frac12)$ & 43 \\ \cline{2-10}
 & $(1,2,5,3) $& &$(\frac{7}{48} ,$&$ \frac16,$&$ -\frac{1}{48} ,$&$ -\frac56 ,$&$ \frac{1}{48},$&$\frac{5}{24})$ & 38 \\ \cline{2-10}
 &$ (2,1,5,2)$ & &$(\frac{1}{12} ,$&$ \frac{7}{24},$&$ -\frac{1}{48} ,$&$ -\frac56 ,$&$ \frac{1}{48}, $&$\frac{5}{24})$ & 44 \\ \cline{2-10}
 & $(3,1,5,2) $& &$(\frac{7}{96} ,$&$ \frac{7}{16},$&$ -\frac{1}{32} ,$&$ -\frac56 ,$&$ \frac{1}{32},$&$\frac{5}{16})$ & 39 \\    \hline 
 \end{tabular}
 \caption{\label{tab:FluxSummary1}The summary of geometric and flux data that lead to three chiral generations in the $\mathbb{Z}_2^{M_2}$ matter parity assignment.
 The first two configurations each contain a redundant flux parameter; we chose to eliminate this redundancy by setting $a_1 =0$.
 For completeness, we have also included the D3-tadpole.}
 \end{center}
 \end{table} 

A few comments about these results are in order.
First, we emphasize that the search procedure is based on the base \textit{independent} flux \eqref{eq:generic_flux} and chirality computation.
This way, we do not have to first construct the full fibration and then construct the fluxes, which, as demonstrated in \cite{Lin:2016vus}, can be very inefficient (largely due to the technical issues of triangulating the polytope of the toric ambient space).
In particular, as our results in Table \ref{tab:FluxSummary1} show, the gauge divisors $W_{2,3}$ are never both toric (i.e., have the divisor class $H$) in fibrations with three generation flux configurations.
Realizing such fibrations explicitly would of course be a good consistency check, but very ineffective for the purpose of scanning a large number of different models.
We content ourselves with the verification that these choices of classes generically do not induce any additional non-abelian gauge divisors or any factorization of the generically present matter curves.
Second, we point out that first two solutions in Table \ref{tab:FluxSummary1} are at the boundary of the allowed region for $(n_2, n_3, s_7, s_9)$, i.e., at these points, some of the coefficients $d_i$ become constant.
Concretely, in the two cases here, it is $d_8$.
When this happens, also the chosen flux basis \eqref{eq:generic_flux} becomes linearly dependent, meaning that one parameter becomes redundant.\footnote{In some cases it can happen, that some constant $d_i$ lead to an enhanced Mordell--Weil group, induced by a multi-section that can become rational. We checked that this jump does not happen in the $[d_8] = 0$ case.}
For concreteness, we set $a_1=0$ in these cases.
Third, note that in all these cases, the D3-tadpole (see below) is positive, which is required for a stable vacuum.
Moreover, the fact that all of them are integer is a necessary condition that fluxes are properly quantized.
In the following, we will provide further arguments for the correct flux quantization in our realizations.

\subsection{Flux quantization and discrete anomalies}

The condition for flux quantizaion reads \cite{Witten:1996md}:
\begin{align}\label{eq:quantization_condition}
	G_4 + \frac{1}{2} c_2(Y) \in H^4(\bbZ, Y) \, ,
\end{align}
where $c_2(Y)$ is the second Chern class of the fourfold.
In practice, verifying this condition explicitly is extremely challenging, and we will not attempt it here.
However, we will perform several non-trivial sanity checks, which in particular involves the relationship of the quantization condition to the gauge anomalies of the $\bbZ_2$.

\subsubsection{D3-tadpole and intersection numbers}

One important quantity which has to be integer for a properly quantized flux is the D3-tadpole \cite{Sethi:1996es},
\begin{align}\label{eq:D3_tadpole}
	n_3 = \frac{\chi}{24} - \frac{1}{2} \int_Y G_4 \wedge G_4 \, .
\end{align}
Here, $\chi$ is the Euler characteristic of the fourfold $Y$, which can be computed as the integral of the fourth Chern class of $Y$ over $Y$.
For toric fibrations, one can compute the Chern classes via adjunction for any base $B$ (see \cite{Klevers:2014bqa, Lin:2015qsa, Lin:2016vus} for examples).
As we already mentioned above, it turns out that the tadpole is always integer for all three generation flux configurations (cf.~table \ref{tab:FluxSummary1}).

Furthermore, given the integrality condition \eqref{eq:quantization_condition}, we necessarily need to have
\begin{align}\label{eq:intersection_with_toric_divisors}
	\left( G_4 + \frac{1}{2} c_2(Y) \right) \cdot D_1 \cdot D_2 \in \bbZ
\end{align}
for any two integer divisor classes $D_1, D_2$.
In practice, we test this condition with $D_i$ being the restrictions of one of the toric divisors from the ambient space.
On the hypersurface, these give rise to the Cartan divisors and the four bisections and are thus manifestly integer classes.
In the above models with three family spectra, all these intersection numbers are integer, thus further supporting the claim that the flux is properly quantized. 

\subsubsection{Geometric incarnation of discrete anomaly cancellation}

One particularly fascinating aspect of the quantization condition is its relationship to the cancellation of discrete anomalies \eqref{eq:discrete_anomaly_cancellation_general}.
Recall that for the case at hand, we are only interested in the $\bbZ_2 - G^2$ anomaly.
For $G = SU(2)$, it receives contributions from doublets which have odd $\bbZ_2$ charge.
Within the spectrum (cf.~Table~\ref{tab:matter_surfaces_charges_top3}) of F-theory on $Y_{31}$, these are the bifundamentals $({\bf 3, 2})$ and the doublets ${\bf 2}_2$.
Therefore, the geometric version of the $\bbZ_2 - SU(2)^2$ anomaly cancellation condition is
\begin{align}
	{\cal A}_{\mathbb{Z}_2-SU(2)^2} = 3 \cdot \chi(({\bf 3,2})) + \chi ( {\bf 2}_2) = G_4 \cdot \left( 3\, [ S_{(\bf 3,2)} ] + [ S_{{\bf 2}_2} ] \right) \in 2\,\bbZ \, .
\end{align}
Inserting the matter surface classes and using the transversality \eqref{eq:transversality_conditions} and gauge symmetry \eqref{eq:gauge_symmetry_condition} constraints of $G_4$, this expression simplifies to
\begin{align}
	{\cal A}_{\mathbb{Z}_2-SU(2)^2}= G_4 \cdot \left( 2\,[y]\,[e_0] - 4\,E_1\,F_1 \right) \, .
\end{align}
Thus, the anomaly is canceled if and only if
\begin{align}\label{eq:Z2_anomaly_geometric_cond}
	{\cal A}_{\mathbb{Z}_2-SU(2)^2} \in 2\,\bbZ \, \Longleftrightarrow \, \frac{1}{2} {\cal A}_{\mathbb{Z}_2-SU(2)^2} = G_4 \cdot \left( [y]\,[e_0] - 2\,E_1\,F_1 \right) \in \bbZ \, .
\end{align}

While this expression depends on the flux and has a priori no reason to be integer, we note that the four cycle class in parenthesis is manifestly integer.
Thus, as long as the flux is properly quantized \eqref{eq:quantization_condition}, one necessarily has to have
\begin{align}
	\left(G_4 + \frac{1}{2} c_2(Y_{31}) \right) \cdot \left( [y]\,[e_0] - 2\,E_1\,F_1 \right) \in \bbZ \, .
\end{align}
Hence, to guarantee \eqref{eq:Z2_anomaly_geometric_cond}, it suffices to show that
\begin{align}\label{eq:Z2_anomaly_chern_class}
	 \frac{1}{2} c_2(Y_{31}) \cdot \left( [y]\,[e_0] - 2\,E_1\,F_1 \right) \in \bbZ \, .
\end{align}
The same method has been used in \cite{Mayrhofer:2014laa} to proof that an F-theory model with $SU(5) \times \bbZ_2$ has no $\bbZ_2$ anomaly.
There, to show that the equivalent version of \eqref{eq:Z2_anomaly_chern_class} held for \textit{any} choice of base and fibration, it was crucial to know how the fluxes and chiralities matched across the conifold transition which unhiggsed the $\bbZ_2$ into a $U(1)$.
Doing the same for our model here is beyond the scope of this paper.
However, we can simply evalute \eqref{eq:Z2_anomaly_chern_class} for every explicit choices of base and fibration structure, in particular on which we found three generation configurations.
And indeed, it turns out that in all fibrations over $\bbP^3$, the $\bbZ_2 - SU(2)^2$ anomaly \eqref{eq:Z2_anomaly_geometric_cond} is canceled due to \eqref{eq:Z2_anomaly_chern_class}.

Likewise, the $\bbZ_2 - SU(3)^2$ anomaly is
\begin{align}
	{\cal A}_{\mathbb{Z}_2-SU(3)^2} = \chi({\bf 3}_1) + \chi ( {\bf 3}_3) = G_4 \cdot \left( [ S_{{\bf 3}_1} ] + [ S_{{\bf 3}_3} ] \right) = G_4 \cdot (2\,F_1\,F_2) \stackrel{!}{\in} 2\,\bbZ \, .
\end{align}
Proceeding analogously as above, this condition is equivalent to
\begin{align}
	\frac{1}{2} c_2(Y_{31}) \cdot F_1 \, F_2 \in \bbZ \, ,
\end{align}
which we can explicitly verify to be true in all cases with $B = \bbP^3$.

\section{Other \texorpdfstring{\boldmath{$\mathbb{Z}_2$}}{Z2} symmetries assignments}
\label{sec:four}
In this section we want to consider other $\mathbb{Z}_2$ charge assignments that can be of phenomenological relevance. These include the other matter parity assignment $\mathbb{Z}_2^{M_1}$ as well as lepton and baryon parity, as discussed in Table~\ref{tab:Z2charges}.
In order to realize them, we consider a different top combination, but perform exactly the same steps that were presented in the previous section.
Hence, we will be brief about the details in this section.

\subsection{Summary of geometric data}

  The top combination that we are considering is $SU(2)$ top 1 and $SU(3)$ top 2, as given in Appendix~\ref{app:B}. 
  The toric data of the model is summarized in Table~\ref{tab:SumTop2}.
\begin{table}[ht]
\renewcommand{\arraystretch}{1.2}
{
\begin{center}
\begin{tabular}{||c|c|c|c|}\hline
 \multicolumn{4}{|c|}{ top vertices \, \, $ \begin{array}{llll } 
SU(2):& e_0:(0,0,1)\, ,& e_1:(1,0,1) \, ,&  \\  
SU(3):& f_0:(0,0,-1)\, ,& f_1:(0,1,-1)\, , & f_2: (-1,1,-1) \end{array}$  }   \\ \hline
 \multicolumn{4}{|c|}{ $ \begin{array}{lllll} 
b_1 = e_0 f_1 f_2^2 d_1 & b_3 = e_0 f_0 f_2d_3 & b_6 = d_6 & b_8 = e_1 f_1 d_8 & b_{10} = e_1f_0^2 f_1 d_{10} \\
b_2 = e_0 f_2 d_2 & b_5= f_1 f_2 d_5 & b_7 = f_0 d_7 & b_9 = e_1f_0 f_1d_9 & 
\end{array}$   }   \\ \hline 
 \multicolumn{4}{|c|}{SRI: \, \, $ \{ x t, x e_1, x f_1, y s, y f_0, t e_0, t f_2, s f_2, e_1 f_2, s f_1 \} \, . $} \\ \hline
 \multicolumn{4}{|c|}{ $\begin{array}{ll} 
 D_{U(1)} =& [x]-[y]-\frac{1}{2}E_1-(\frac{1}{3}F_1+\frac{2}{3}F_2)-\frac{1}{3}W_3+\frac{1}{2}\overline{K}_B+\frac{1}{2}S_7-\frac{1}{2}S_9\, , \\
 D_{\mathbb{Z}_2}=& x + \frac23 (F_1 + 2 F_2) \, .
\end{array}$} \\ \hline 

 \multicolumn{4}{|c|}{  $  \begin{tabular}{lll} 
    $ [d_1] = 3\overline{K}_B - S_7 - S_9 - W_2 $ &$ [d_2] = 2\overline{K}_B-S_9  - W_2$  &     
$[d_3] = \overline{K}_B + S_7 - S_9 - W_3-W_2$  \\ 
 $ [d_5] = 2\overline{K}_B - S_7 $&$ [d_6] = \overline{K}_B$ &$ [d_7] = S_7 - W_3$ \\
$ [d_8] =\overline{ K}_B + S_9 - S_7 $&$ [ d_9 ] = S_9 - W_3 $& 
$  [d_{10} ] = S_9 + S_7 - \overline{K}_B - 2 W_3$ 
 \end{tabular} $ } \\ \hline 
 \end{tabular}
 \end{center}}
\caption{\label{tab:SumTop2} Geometric data for the hypersurface specialization of the second top combination. }
\end{table}
%From the divisor classes that specify this top combination it becomes clear, that this model is in fact equivalent to the top combination 1 that was studied in Section~\ref{sec:three}. To see this
%explicitly, one has to perform the shift of divisor classes as $ S_9 \rightarrow S_9 + W_3$.\TODO{This is simply wrong.}
%This already hints at the fact the two top combinations are geometrically equivalent and related by an $GL(4,\mathbb{Z})$ transformation.  
The matter loci can easily be determined by the information of the two tops given in Appendix~\ref{app:B} including the additional bifundamental representation at $W_2 = W_3 =0$  as summarized in Table~\ref{tab:matter_surfaces_charges2_top2}. 
The spectrum is very similar as before but includes a $\bbZ_2$-even charged bifundamental.
\begin{table}[ht]
\begin{align*}
	\begin{array}{|>{ \displaystyle}  c |  >{\displaystyle} c | >{\displaystyle} c | >{\displaystyle} c| >{\displaystyle} c|} \hline
		\text{Label} & G_{\text{SM}} \times \mathbb{Z}_2 \text{ Rep.} &  \mathbb{Z}_2^{M_1} & \mathbb{Z}_2^L&\mathbb{Z}_2^B \\ \hline \hline \rule{0pt}{4ex}
		{\bf 2}_1 & (\mathbf{1},\mathbf{2})_{(-\frac12,-)} & H_d , H_u & L & L\\ \rule{0pt}{4ex}
		{\bf 2}_2 & (\mathbf{1},\mathbf{2})_{(-\frac12,+)} & L & H_d, H_u & - \\ \rule{0pt}{4ex}
		\overline{\bf 3}_1 &  (\overline{\mathbf{3}},\mathbf{1})_{(-\frac23,-)}  & \overline{u} & -& \overline{u}  \\ \rule{0pt}{4ex} 
		\overline{{\bf 3}}_2 &   (\overline{\mathbf{3}},\mathbf{1})_{(-\frac23,+)}  & - & \overline{u} & - \\ \rule{0pt}{4ex}
	{\bf 3}_3 &({\mathbf{3}},\mathbf{1})_{(-\frac13,-)}  &\overline{d} & - & \overline{d}  \\ \rule{0pt}{4ex}
		{\bf 3}_4 &({\mathbf{3}},\mathbf{1})_{(-\frac13,+)} & - & \overline{d} & -  \\ \rule{0pt}{4ex}
		{\bf (\overline{3},2) } &( \overline{\mathbf{3}} ,\mathbf{2})_{(-\frac16,+)} & Q & Q & Q \\ \rule{0pt}{4ex}
		{\bf   \mathbf{1}  }_1 & (\mathbf{1},\mathbf{1})_{(1,-)}  & \overline{e} & \overline{e} & - \\ \rule{0pt}{4ex} 
	{\bf   \mathbf{1}  }_2 &  (\mathbf{1},\mathbf{1})_{(-1,+)}  & -& -&   \overline{e} \\  \hline  
	\end{array}
\end{align*}
\caption{\label{tab:matter_surfaces_charges2_top2} Matter curves and their charges as given for the second top combination. 
MSSM field identifications under various $\mathbb{Z}_2$ symmetries are given in the last three columns. 
For each identification we assign the chirality three to MSSM fields whereas states marked with a ``$-$'' must be non-chiral.}
\end{table}
This allows for a straightforward identification of the geometric $\bbZ_2$ with the other three parities listed in Table \ref{tab:Z2charges}.

%Although the two top combinations are equivalent, the positive $\mathbb{Z}_2$ charge for the bifundamental makes the charge assignments more evident (compare with Table~\ref{tab:Z2charges}).
 The matter homology classes of the curves in terms of ambient divisors is given in Table~\ref{tab:top2x1_matter_homology_class} of Appendix~\ref{app:A} which can be used to obtain the five independent matter surface fluxes:
 \begin{equation}
\begin{split}
 A({\two_{2}})&=[S_{\two_2}]-\frac{1}{2}E_1 \cdot (-2W_2-W_3+6\overline{K}_B-2S_9) \, , \\
A({\overline{\three}_2}) &=[S_{\three_2}]-\frac{1}{3}(2F_1+F_2) \cdot (2\overline{K}_B-W_2-W_3+S_7-S_9)+\frac{1}{3}C_{\three_2}\, ,\\
A({\three}_{4})&=[S_{{\three_4}}]+\frac{1}{3}(2F_1+F_2)\cdot (-W_2+5\overline{K}_B-S_7-S_9)-\frac{1}{3}C_{\three_4} \, , \\
A(({\three},\two)&=[S_{({\three},\two)}]+\frac{1}{3}(-F_1+F_2)\cdot W_2+\frac{1}{2}E_1\cdot W_3 -\frac{1}{3}W_2\cdot W_3\, , \\
A(\overline{\one}_2)&=[S_{\overline{\mathbf{1}}_2}] \, ,
\end{split}
\end{equation}
where $C_{\bf R}$ denotes the classes of the associated matter curves.
These admit the following algebraic equivalence relations between the above flux basis and other vertical 4-cycles: 
\begin{equation}
    \begin{split}
        A(\two_2)-A(({\three},\two))-2D_{U(1)}\cdot W_2+A(\two_1) & \, =0 \, ,\\
        A((\overline{\three},\two))-A(\overline{\three}_2)+D_{U(1)}\cdot W_3-A(\overline{\three}_1) & \, =0 \, ,\\
        A({\three}_4)+A(({\three},\two))-D_{U(1)}\cdot W_3+A({\three}_3) & \, =0 \, ,\\
        A(({\three},\two))+A(\one_{2})+D_{U(1)}\cdot (-6\overline{K}_B+2W_2+3W_3)-A(\one_{1}) & \, = 0 \, .
    \end{split}
\end{equation} 
The $G_4$-flux in the above basis is then given by
\begin{equation}
    G_4=a_1A(\two_2)+a_2A(\three_2)+a_3A(\three_4)+a_4A((\overline{\three},\two))+a_5A(\overline{\one}_{2})+D_{U(1)}\wedge F \, .
\end{equation}

\subsection{Three family searches and discrete anomalies}

For concrete three family realizations, we again pick the base to be $\bbP^3$.
In this model, we now have the possibility to assign different physical interpretations to the $\bbZ_2$.
In each fibration parametrized by $(n_2,n_3, s_7, s_9)$, we search for flux configurations $(a_1, a_2, a_3, a_4, a_5, \lambda)$ compatible with one of the three possible identifications listed in Table \ref{tab:matter_surfaces_charges2_top2}.
In all of them, we again impose the vanishing of the fluxed induced D-term of the $U(1)$.
Those solutions that also have a positive integer D3-tadpole are listed in Table~\ref{tab:FluxSummary2}.
\begin{table}[ht]
 \begin{center}
 \renewcommand{\arraystretch}{1.4}
 \begin{tabular}{|c|c|c c c c c c |c|}\hline
 & $(n_2, n_3, s_7, s_9)$ & $(a_1,$&$a_2,$&$a_3,$&$a_4,$&$a_5,$&$\lambda)$ & $n_{D3}$ \\ \hline
 \multirow{5}*{$\mathbb{Z}_2^{M_1}$} 
& (1,3,7,4) & $(0,$&$-\frac74,$&$ \frac34,$&$2,$&$\frac34,$&$-\frac32)$ & 33 \\ \cline{2-9}
& (3,1,5,4) & $(0,$&$-\frac{5}{12},$&$ \frac{1}{12} ,$&$\frac23 ,$&$  \frac{1}{12},$&$-\frac12) $& 43 \\ \cline{2-9}
& (1,2,5,4) & $(\frac{7}{48} , $&$-\frac16,$&$ \frac{1}{48} ,$&$ -\frac{11}{16} ,$&$ \frac{1}{48},$&$-\frac{1}{12})$ & 38 \\ \cline{2-9}
& (2,1,5,4) & $(\frac{1}{12} ,$&$ -\frac{7}{24},$&$ \frac{1}{48} ,$&$ \frac34 ,$&$ \frac{1}{48},$&$-\frac{1}{8})$ & 44 \\ \cline{2-9}
& (3,1,5,3) & $(\frac{7}{96} ,$&$ -\frac{7}{16},$&$ \frac{1}{32} ,$&$ -\frac{73}{96} ,$&$ \frac{1}{32},$&$-\frac{1}{8})$ & 39 \\    \hline
 \multirow{5}*{$\mathbb{Z}_2^{L}$}
 &(1,2,4,4) &  $(0,$&$ \frac{5}{16},$&$-\frac{5}{96} ,$&$  \frac{15}{32} ,$&$  \frac16 ,$&$  -\frac{73}{48} )$  & 43\\\cline{2-9}
 &(1,2,5,3) &$(0,$&$ \frac{1}{12}  ,$&$-\frac{1}{12} ,$&$  \frac12 ,$&$   0,$&$ 0)$  & 42\\ \cline{2-9}
 &(1,2,5,5) &$(0,$&$\frac14,$&$-\frac14,$&$\frac12,$&$0,$&$0)$  & 40\\ \cline{2-9}
 &(3,2,5,3) &$(0,$&$ \frac14 ,$&$-\frac14,$&$  \frac16,$&$  0,$&$  0)$  & 32\\ \cline{2-9}
 &(1,2,6,4) &$(\frac{41}{384},$&$  \frac{7}{32},$&$  -\frac{61}{384}, $&$\frac{175}{384},$&$  -\frac{23}{384},$&$ \frac{23}{96})$  & 39\\  \hline
 \multirow{2}*{$\mathbb{Z}_2^{B}$}
 &(3,2,5,4) &$(0,$&$ 0,$&$-\frac13,$&$  \frac13 ,$&$  -\frac16 ,$&$  \frac53 )$  & 34\\\cline{2-9}
 &(1,2,5,4) &$(-\frac{1}{12},$&$-\frac13,$&$-\frac{1}{12},$&$\frac34,$&$-\frac{1}{12},$&$\frac43)$  & 38\\  \hline
 \end{tabular}
 \caption{\label{tab:FluxSummary2}The summary of geometric and flux quanta that lead the three chiral generations for the three discrete symmetry assignments of the second top combination.      }
 \end{center}
 \end{table}
 First we note that the solutions we obtain for the second matter parity assignment have a very similar structure compared to the models we obtained in the previous section (cf.~Table \ref{tab:FluxSummary1}), including the same number of D3-branes.
 This points towards an equivalence between the two fibrations defined via the two different top combinations.\footnote{The classes of the sections of the two tops can be related by a change $d_1 \leftrightarrow d_8, d_9 \leftrightarrow d_2, d_{10} \leftrightarrow d_3$.}
 A more general analysis of this equivalence including the necessary redefinitions of the abelian symmetry generators is left for future research.
% First we note that the solutions for the second matter parity, are precisely\TODO{The only additional data that is the same is D3-tadpole. This hardly counts as `precisely the same'} the same ones 
% as we found in the first top combination, after applying the shift in divisor classes. This confirms that the field theoretic equivalent charge assignments are also geometrically equivalent.\TODO{Again, not true.}
 Secondly it is important to emphasize that only the flux configurations for matter parity fulfill the $G_4$-flux integrality condition 
\begin{align}
\label{eq:SU2Anomaly2}
(G_4 + \frac12c_2(Y) )\cdot D_1 \cdot D_2 \in \mathbb{Z} \, ,
\end{align}
whereas the lepton and baryon parity assignments do not. 
Again, we can nicely relate this rather obscure geometric property directly to the cancellation of $\bbZ_2$ anomalies.

Namely, these are
\begin{align} 
\begin{split}
	\mathcal{A}_{Z_2-SU(2)^2}= \, & \chi(\mathbf{2}_1) = G_4 \cdot ( [S_{\mathbf{2}_1}]) = -2 \, [y]\,E_0 \in 2 \mathbb{Z} \, \\
	\mathcal{A}_{Z_2-SU(3)^2}= \, & \chi(\mathbf{3}_1)+\chi(\mathbf{3}_3) = G_4 \cdot ( -[S_{\overline{\mathbf{3}}_1}] + [S_{\mathbf{3}_3}]) = 2\,F_1 \, F_2 \in 2 \mathbb{Z} \, .
\end{split}
\end{align}
Like in the previous section, the quantization condition translates these conditions into a question about integrality of the intersection numbers
\begin{align}
	\frac12 c_2 (Y_{32}) \cdot \left\{\begin{array}{>{\displaystyle} c} [y]\,E_0 \\ F_1\,F_2 \end{array} \right. \, ,
\end{align}
which both turn out to be indeed integral for all the fibrations we scanned over.
However, we also know that only the matter parity $\bbZ_2^{M_1}$ assignment of the chiralities is anomaly free, whereas the lepton and baryon parity assignments are not with the spectrum in \ref{tab:matter_surfaces_charges2_top2}.
Since the flux configurations are chosen to reproduce these chiral spectra, we arrive at the same conclusion---but based on field theoretic anomaly considerations---that the flux solutions for these two assignments in Table \ref{tab:FluxSummary2} cannot be properly quantized.

\section{Summary and Conclusion}
\label{sec:five}
In this work we have engineered globally consistent four dimensional MSSM-like particle physics models with three chiral generations that admit $\mathbb{Z}_2$ quantum numbers under the matter parity extension of the Standard Model gauge group.
Our compactifiations are genus-one fibered fourfolds with $G_4$-flux over a simple $\mathbb{P}^3$ base space that pass all necessary consistency conditions:
The $G_4$-flux is properly quantized, and the D3-tadpole is canceled with a positive and integral number of D3-branes. 
For this explicit construction we employed toric geometry to engineer the resolved fourfold which allows the direct computation of all (discrete) gauged quantum numbers. 
In addition, the fact that the internal space is smooth allows us to easily handle the gauge background, giving us the power to scan systematically for configurations leading to three chiral generations.
These constructions are flexible enough to also allow, at least in principle, for other $\mathbb{Z}_2$ symmetries, such as lepton and baryon parity, by choosing different flux configurations. 
These models however suffer from non-properly quantized $G_4$-fluxes, even though they give three chiral families and a positive integer number of D3-branes. 
We have shown that this is not just a coincidence, but actually intimately related to the fact that lepton and baryon parities are not free of discrete anomalies with just the MSSM spectrum.

However several interesting questions remain. 
First it would be exciting to investigate the interplay between fluxes and discrete anomalies further. 
For example, an analysis similar to \cite{Mayrhofer:2014haa, Lin:2015qsa} of the conifold transition that unhiggses the $\bbZ_2$ into a $U(1)$ could allow us to proof the cancellation of discrete anomalies for \textit{generic} fibrations.
Moreover we have left out possible Abelian-$\mathbb{Z}_2$ anomalies as they are hard to investigate in the field theory due to an ambiguous $U(1)$ charge normalization \cite{Ibanez:1991hv}. However, since in F-theory there is a natural charge quantization inherited from the Mordell--Weil lattice \cite{Cvetic:2017epq}, one might hope that a more geometric treatment of this issue is possible.  
Further important steps towards more realistic phenomenology is to understand the full vector-like sector and to decouple possible (vector-like) exotics while keeping one pair of Higgs doublets light. 
This would also allow us to determine the presence of right-handed neutrinos, whose representation is, at least in principle, realized explicitly in the geometry.
Due to recent progress \cite{Bies:2014sra,Bies:2017fam} this goal seems to be in reach.
However, applying the methods presented there to a complex configuration of matter curves, such as we have in our models, are not feasible with the given algorithms and computing power today.
But even without exotics, this model might still suffer from higher dimensional operators in the effective action such as
\begin{align}
\mathcal{W} \ni \kappa^1 QQQL +  \kappa^2 \overline{u} \overline{u}\overline{d}  E \, ,
\end{align}
whose coefficients are strongly constrained by proton decay but can not be forbidden with matter parity alone. Hence one might want to construct higher order discrete symmetries, ideally the $\mathbb{Z}_6$ proton hexality \cite{Dreiner:2005rd} which forbids also other dangerous higher dimensional operators, and is anomaly free. The classification or construction of higher order (possibly non-Abelian \cite{Grimm:2015ona,Braun:2017oak,Cvetic:2018xaq}) discrete symmetries base-independently beyond $\mathbb{Z}_4$ \cite{Braun:2014qka, Oehlmann:2016wsb} are unknown yet (see, however, \cite{Kimura:2016crs} for some recent examples over specific bases) and, hence, a topic of great interest. Once such a classification is available, we hope that a generalization of our work can realize the chiral MSSM with such a discrete symmetry extension. 

\section*{Acknowledgments}
We thank Lara Anderson, James Gray, Markus Dierigl and Timo Weigand for discussions and comments. 
We further thank Denis Klevers and Jonas Reuter for contributions during the early stages of this project. 
P.-K.~O. is grateful to the Yau Mathematical Sciences Center for its hospitality during the completion of this work. 
M.~C., M.~L. and L.~L. are supported by DOE Award DE-SC0013528. M.C.~further
acknowledges the support by the Fay R.~and Eugene L.~Langberg Endowed Chair and
the Slovenian Research Agency The work of P.K.O.~is supported by an individual DFG grant OE 657/1-1.

\appendix

\section{Homology classes of matter surfaces}
\label{app:A}

In this appendix, we collect the matter surface homology classes of the two top combinations considered in Sections \ref{sec:three} and \ref{sec:four}, in Table \ref{tab:top3x1_matter_homology_class} and \ref{tab:top2x1_matter_homology_class}, respectively.
\begin{table}[ht]
\begin{center}
{\small
\renewcommand{\arraystretch}{1.1}
\begin{tabular}{|@{}c@{}|c|}\hline
$\bf R$ & Matter surface homology classes \\ \hline

\multirow{ 2}{*}{$\two_1$} & $-(E_1F_1) - E_1\overline{K}_B + E_1S_7 - E_1S_9 + F_1W_2 + \overline{K}_BW_2 - S_7W_2 + S_9W_2 - W_2[x]$\\ 
& $ - W_2[x] - 2E_1[y] + 2W_2[y]
$ \\ \hline

\multirow{ 1}{*}{$\two_2$} & $ 3E_1\overline{K}_B - E_1S_7 - E_1S_9 + F_2W_2 - \overline{K}_BW_2 + S_9W_2 - W_2[x] + 2E_1[y] $\\
 \hline

$\overline{\three}_1$& $F_2^2 - F_2\overline{K}_B + F_2S_9 + F_2W_2 - E_1W_3 - F_2W_3 + \overline{K}_BW_3 - S_9W_3 + W_3[x]
$ \\ \hline

${\three}_2$& $-(E_1F_1) + F_2^2 + F_1\overline{K}_B + F_2\overline{K}_B - F_2S_7 + W_3[y]
$ \\ \hline

$\overline{\three}_3$& $-2F_1F_2 - F_2^2 + F_2S_7
$ \\ \hline

$\overline{\three}_4$& $E_1F_1 + 2F_1F_2 + F_2^2 - 2F_1\overline{K}_B - F_2\overline{K}_B - F_2S_9 + F_1W_3 + 2\overline{K}_BW_3$\\
&$  - S_7W_3 + S_9W_3 - W_3[x] + W_3[y]
$ \\ \hline

$(\overline{\three},\two) $& $E_1F_0$\\\hline

\multirow{ 3}{*}{${\one}_{1}$} & $-2F_1F_2 - F_2^2 + E_1\overline{K}_B + 3F_1\overline{K}_B + F_2\overline{K}_B + 2\overline{K}_B^2 - E_1S_7 - F_1S_7 - 3\overline{K}_BS_7 + S_7^2 $ \\
& $ + E_1S_9- F_1S_9 + F_2S_9 + 2\overline{K}_BS_9 - S_7S_9 - F_1W_2 - \overline{K}_BW_2 + S_7W_2 - S_9W_2 $ \\
& $ - 4\overline{K}_B[x] + 2S_7[x] + W_2[x] + 2E_1[y] + 2\overline{K}_B[y] + 2S_9[y] - 2W_2[y] - 2W_3[y] - 4[x][y]$ \\ \hline

\multirow{ 4}{*}{${\overline{\one}}_{2}$} & $-(E_1F_1) - 2F_1F_2 - F_2^2 - 2E_1\overline{K}_B + F_1\overline{K}_B - 3F_2\overline{K}_B + 2\overline{K}_B^2 - E_1S_7 - F_1S_7$ \\
& $ + 2\overline{K}_BS_7 + E_1S_9 - F_1S_9 + F_2S_9 - 3\overline{K}_BS_9 - S_7S_9 + S_9^2 + E_1W_2 + F_2W_2$ \\
& $  - \overline{K}_BW_2+ 2E_1W_3 + F_1W_3 + 2F_2W_3 - 3\overline{K}_BW_3 + 2S_9W_3 + 2\overline{K}_B[x] + 2S_7[x]$ \\
& $  - W_2[x] - 3W_3[x] + 2E_1[y]- 4\overline{K}_B[y] + 2S_9[y] + W_3[y] - 4[x][y] $ \\ \hline

\end{tabular}
}
\caption{Summary of matter homology classes restricted to CY of first top combination.}\label{tab:top3x1_matter_homology_class}

\end{center}
\end{table}

\clearpage

\begin{table}[ht]
\begin{center}

{\small
\renewcommand{\arraystretch}{1.1}
\begin{tabular}{|@{}c@{}|c|}\hline
${\bf R}$ & Matter surface homology classes \\ \hline

\multirow{ 2}{*}{$\two_1$} & $-(E_1\overline{K}_B) - E_1S_7 - E_1S_9 + \overline{K}_BW_2 + S_7W_2 + S_9W_2 $\\ 
& $+ 2E_1W_3 - 2W_2W_3 + W_2[x] + 2E_1[y] - 2W_2[y]
$ \\ \hline

\multirow{ 2}{*}{$\two_2$} & $-(E_1F_1) + 3E_1\overline{K}_B + E_1S_7 - E_1S_9 - 2E_1W_2 + F_2W_2 $\\
& $+ \overline{K}_BW_2 - S_9W_2 - E_1W_3 + W_2[x] - 2E_1[y]
$ \\ \hline

$\overline{\three}_1$& $F_2^2 + F_2\overline{K}_B - F_2S_9 + W_3[x]
$ \\ \hline

$\overline{{\three}}_2$& $-(E_1F_1) - F_2^2 + F_1\overline{K}_B - F_2\overline{K}_B + F_1S_7 + F_2S_7 - F_1S_9 - F_1W_3 - W_3[y]
$ \\ \hline

${\three}_3$& $2F_1F_2 + F_2^2 + F_2\overline{K}_B + F_2S_9 - 2F_2W_3
$ \\ \hline

${\three}_4$& $E_1F_1 - 2F_1F_2 - F_2^2 - 3F_1\overline{K}_B - 2F_2\overline{K}_B + F_1S_7+ F_2S_7 + F_1S_9$\\
&$  - E_1W_3 - F_1W_3 + F_2W_3 + 2\overline{K}_BW_3 - S_9W_3 + W_3[x] - W_3[y]
$ \\ \hline

$(\overline{\three},\two) $& $-(E_1F_1) + F_1W_2$\\\hline

\multirow{ 3}{*}{${\one}_{1}$} & $2F_1F_2 + 3F_2^2 + E_1\overline{K}_B + F_1\overline{K}_B + 3F_2\overline{K}_B + 2\overline{K}_B^2 - E_1S_7 - F_1S_7 - 2F_2S_7 - 3\overline{K}_BS_7 $ \\
& $ + S_7^2 + E_1S_9 + F_1S_9 - F_2S_9 + 2\overline{K}_BS_9 - S_7S_9 - \overline{K}_BW_2 + S_7W_2 - S_9W_2 - 4\overline{K}_B[x] $ \\
& $ + 2S_7[x] + W_2[x] + 2E_1[y] + 2\overline{K}_B[y] + 2S_9[y] - 2W_2[y] - 2W_3[y] - 4[x][y]$ \\ \hline

\multirow{ 4}{*}{${\overline{\one}}_{2}$} & $E_1F_1 + 2F_1F_2 + 3F_2^2 - 2E_1\overline{K}_B - F_1\overline{K}_B + 5F_2\overline{K}_B + 2\overline{K}_B^2 - E_1S_7 - F_1S_7 - 2F_2S_7$ \\
& $ + 2\overline{K}_BS_7 + E_1S_9 + F_1S_9 - F_2S_9 - 3\overline{K}_BS_9 - S_7S_9 + S_9^2 + E_1W_2 - F_2W_2 - \overline{K}_BW_2$ \\
& $ + S_9W_2 + E_1W_3 + F_1W_3 - F_2W_3 - 2\overline{K}_BW_3 + S_9W_3 + 2\overline{K}_B[x] + 2S_7[x] - W_2[x] $ \\
& $ - 3W_3[x]+ 2E_1[y] - 4\overline{K}_B[y] + 2S_9[y] + W_3[y] - 4[x][y] $ \\ \hline

\end{tabular}
}
\caption{Matter surface homology classes of the second top restricted on the fourfold.}\label{tab:top2x1_matter_homology_class}
\end{center}
\end{table}

\section{Towards the vector-like spectrum of the first top}
\label{app:C}
In this section we want to give all information that is needed to compute the vector-like spectrum of our first model realizing $\bbZ_2^{M_2}$ using the methods of \cite{Bies:2017fam}. 
The key point is to assign to each matter curve $C_{\bf R} \subset B$ a divisor $D$, i.e., a collection of points on $C_{\bf R}$, based on the intersection properties between the $G_4$-flux and the matter surface $S_{\bf R}$.
By expression the flux in terms of matter surfaces, evaluating this intersection product reduce to properly counting the points, in which various matter surfaces meet.
In F-theory geometries, these points are in the fibers over codimension three enhancement loci, i.e., Yukawa points $Y_i$.
Thus, the resulting divisor $D$ is a linear combination $\sum_i \mu_i Y_i$ of these points.
From this divisor, one can then extract the left- and right-handed fermions as the sheaf cohomologies
\begin{align}
	h^i ( C_{\bf R} \,, \,  \cO_{C_{\bf R}}(D) \otimes \sqrt{K_{C_{\bf R}}} ) \, , \quad i = 0,1 \,  ,
\end{align}
where $\sqrt{K_{C_{\bf R}}}$ is the spin bundle on $C_{\bf R}$.
These cohomologies depend on the complex structure parameters of the fourfold.
However, the chiral index $\chi = h^0 - h^1$ is a topological invariant, which is simply the number of points (including signs) that constitutes $D \subset B$.
More details and examples can be found in \cite{Bies:2017fam}.

For the first top realizing the MSSM with matter parity $\bbZ_2^{M_2}$, we need, in addition to the particular flux basis \eqref{eq:generic_flux} that we have picked, also algebraic equivalence relations between fluxes and other vertical 4-cycles.
Explicitly, these are
\begin{align}
	\begin{split}
		A({\bf 2}_2) - A( {\bf (3,2)}) + 2\,D_{U(1)} \cdot W_2 + [S_{{\bf 2}_1}] + \frac{1}{2}\,E_1 \cdot [\{f_{{\bf 2}_1}\}] - \frac{1}{2}\,C_{{\bf 2}_1} = 0 & \, , \\
		A( {\bf (3,2)})  - A({\bf 3}_2) - D_{U(1)} \cdot W_3 + [S_{{\bf 3}_1}]  + \frac{1}{3}\,(F_1 + 2\,F_2) \cdot [\{f_{{\bf 3}_1} \}] - \frac{1}{6}\,C_{{\bf 3}_1} = 0 & \, , \\
		A({\bf 3}_4) + A( {\bf (3,2)}) + D_{U(1)} \cdot W_3 + [S_{{\bf 3}_3}] + \frac{1}{3}\,(F_1 - F_2) \cdot [\{ f_{{\bf 3}_3} \}] - \frac{1}{6}\, C_{{\bf 3}_3} = 0 & \, , \\
		A({\bf (3,2)}) -A({\bf 1}_{(1,+)}) + D_{U(1)} \cdot (6\,\KB - 2\,W_2 - 3\,W_3) + [S_{{\bf 1}_{(1,-)}}] - \frac{1}{2} \, C_{{\bf 1}_{(1,-)}} = 0 & \, .
	\end{split}
\end{align}

\begin{table}[ht]
\begin{center}
{\small
\renewcommand{\arraystretch}{1.2}
\begin{tabular}{|c|c|}\hline

Coupling  &Homology classes \\ \hline

\multirow{2}{*}{$\two_1\cdot \two_2 \cdot \overline{\one_{(1,-)}}$}  
 &  $[Y_1]=W_2\cdot (6\overline{K}_B^2 - 4\overline{K}_BS_7 + 2S_7^2 + 4\overline{K}_BS_9 - 2S_9^2  - 2\overline{K}_BW_2$  \\ 
      & $ - 2S_9W_2  - \overline{K}_BW_3 - 3S_7W_3 - S_9W_3 + W_2W_3 + 2W_3^2)$\\     \hline
$\two_1 \cdot \two_1 \cdot \overline{\one_{(1,+)}}$  & $[Y_2]=W_2\cdot (2\overline{K}_BS_7 - S_7^2 + \overline{K}_BS_9 + S_9^2 - 2\overline{K}_BW_3 + S_7W_3 - S_9W_3)$
 \\ \hline
\multirow{2}{*}{$\two_1 \cdot \overline{\two_2} \cdot \one_{(0,-)}$} &    $[Y_3]=W_2\cdot (6\overline{K}_B^2 + 4\overline{K}_BS_7 - 2S_7^2 + 4\overline{K}_BS_9 - 2S_9^2 - 2\overline{K}_BW_2$ \\
&  $- 2S_9W_2  - 9\overline{K}_BW_3 + 3S_7W_3 - S_9W_3 + W_2W_3) $ \\ \hline
\multirow{2}{*}{$\two_2 \cdot \two_2 \cdot \overline{\one_{(1,+)}}$} & $[Y_4]=
W_2 \cdot (6\overline{K}_B^2 + 2\overline{K}_BS_7 - S_7^2 - 5\overline{K}_BS_9 + S_9^2 - 5\overline{K}_BW_2 + 2S_9W_2 $ \\
 &  $+ W_2^2  - 7\overline{K}_BW_3 + 2S_7W_3 + 2S_9W_3 + 2W_2W_3)$ \\ \hline

$\two_1 \cdot {\three}_{4} \cdot (\overline{\three},\two)$ & $[Y_5]=W_2\cdot W_3 \cdot (2\overline{K}_B - S_7 + S_9)$ \\ \hline

$\two_1 \cdot \overline{\three_{2}} \cdot ({{\three},\two})$  & $[Y_6]=W_2\cdot W_3 \cdot (S_7 + S_9 - W_3)$ \\ \hline
$\two_2 \cdot {\three}_{3} \cdot (\overline{\three},\two)$  & $[Y_7]=W_2\cdot W_3 \cdot (3\overline{K}_B + S_7 - S_9 - W_2 - 2W_3)
$\\ \hline
$\two_2 \cdot \overline \three_{1} \cdot ({{\three},\two})$  & $[Y_8]=
W_2\cdot W_3 \cdot (3\overline{K}_B - S_7 - S_9 - W_2)$ \\ \hline
$\overline{\three}_1 \cdot \overline{\three}_3 \cdot \overline{\three}_4 $  & $[Y_9]=
W_3 \cdot \overline{K}_B(3\overline{K}_B - S_7 - S_9 - W_2)$ \\ \hline
$\three_2 \cdot {\three}_3 \cdot {\three}_3 $  & $[Y_{10}]=W_3 \cdot \overline{K}_B(S_7 - W_3)$ \\ \hline
$\three_2 \cdot {\three}_4 \cdot {\three}_4 $  & $[Y_{11}]=
W_3 \cdot \overline{K}_BS_9$  \\ \hline
$\overline{\three}_1 \cdot {\three}_3 \cdot \one_{(1,+)} $  & $[Y_{12}]=
W_3 \cdot (3\overline{K}_B - S_7 - S_9 - W_2)(2\overline{K}_B + S_7 - S_9 - W_2 - 2W_3)$ \\ \hline
$\overline{\three}_1 \cdot {\three}_4 \cdot \one_{(1,-)} $  & $[Y_{13}]=
W_3 \cdot (2\overline{K}_B - S_7 + S_9)(3\overline{K}_B - S_7 - S_9 - W_2)$ \\ \hline
\multirow{2}{*}{$\three_2 \cdot \overline{\three}_3 \cdot \overline{\one_{(1,-)}} $}  &  $[Y_{14}]=W_3 \cdot (\overline{K}_BS_7 + S_7^2 + 3\overline{K}_BS_9 - S_9^2 - S_7W_2 - S_9W_2- \overline{K}_BW_3 $\\ & $ - 3S_7W_3 - S_9W_3 + W_2W_3 + 2W_3^2)$\\ \hline
$\three_2 \cdot \overline{\three}_4 \cdot \overline{\one_{(1,+)}} $ & $[Y_{15}]=
W_3 \cdot (3\overline{K}_BS_7 - S_7^2 + \overline{K}_BS_9 + S_9^2 - 3\overline{K}_BW_3 + S_7W_3 - S_9W_3)$\\ \hline
$\overline{\three}_1 \cdot \three_2 \cdot \one_{(0,-)} $  & $[Y_{16}]=W_3 \cdot (3\overline{K}_B - S_7 - S_9 - W_2)(S_7 + S_9 - W_3)$ \\ \hline
\multirow{2}{*}{$\overline{\three}_3 \cdot {\three}_4 \cdot \one_{(0,-)} $} &  $[Y_{17}]=
W_3\cdot (6\overline{K}_B^2 + \overline{K}_BS_7 - S_7^2 + \overline{K}_BS_9 + 2S_7S_9 - S_9^2 - 2\overline{K}_BW_2$\\
& $ + S_7W_2 - S_9W_2 - 6\overline{K}_BW_3 + 2S_7W_3 - 2S_9W_3)$ \\ \hline
$(\overline{\three},\two)$ $ \cdot (\overline{\three},\two) \cdot \overline{\three}_4 $  & $[Y_{18}]=
W_2\cdot W_3 \cdot \overline{K}_B$ \\ \hline

\end{tabular}
}
\caption{\label{tab:poly11_yukawa}Yukawa points of the first top.}
\end{center}
\end{table}

The Yukawa points and their homology classes are listed in Table \ref{tab:poly11_yukawa}.
In terms of the homology classes, we can write the chiral indices of the matter states induced by the flux \eqref{eq:generic_flux} as
\begin{align}\label{eq:chiralities_top1}
    \begin{split}
        \chi(\overline{\three}_1)=&-\frac{2}{3}C_{\overline{\three}_1} \,  F + a_1[Y_8]+\frac{1}{3}a_2[Y_{16}]+a_3\, (\frac{2}{3}\,  [Y_9]-\frac{1}{3}\,  [Y_{13}])-\frac{1}{3} a_4[Y_8]-a_5[Y_{12}] \, ,\\
        \chi(\three_2)=&\frac{2}{3}C_{\three_2}\, F+ a_2\,  (\frac{1}{3}\,  [Y_{16}]+\frac{1}{3}\,  [Y_6]-\frac{2}{3}\,  C_{\three_2}\,  W_3)-a_3(\frac{4}{3}[Y_{11}]-\frac{1}{3}[Y_{15}]) \\
        & +\frac{1}{3}a_4[Y_6]+a_5[Y_{15}] \, ,\\
        \chi(\overline{\three}_3)=&\frac{1}{3}C_{\overline{\three}_3}\, F-a_1[Y_7]+a_2(-\frac{4}{3}[Y_{10}]+\frac{1}{3}[Y_{14}])+a_3(\frac{2}{3}[Y_9]-\frac{1}{3}[Y_{17}])-\frac{1}{3}a_4[Y_7]+a_5[Y_{12}] \, ,\\
        \chi(\overline{\three}_4)=&\frac{1}{3}C_{\overline{\three}_4}\, F+a_2(-\frac{4}{3}[Y_{11}]+\frac{1}{3}[Y_{15}])+a_3(\frac{1}{3}[Y_5]-\frac{2}{3}[Y_{18}]-\frac{2}{3}[Y_{9}]+\frac{1}{3}[Y_{17}]-\frac{1}{3}C_{\overline{\three}_4}W_3)\\
        &+a_4(-\frac{1}{3}[Y_5]+\frac{2}{3}[Y_{18}])+a_5[Y_{15}] \, , \\
        \chi(\overline{\three},\two)=&-\frac{1}{6}C_{(\overline{\three},\two)}\, F+\frac{1}{2}a_1([Y_7]-[Y_8])+\frac{1}{3}a_2[Y_6]+\frac{1}{6}a_4C_{(\overline{\three},\two)}\,  (6\overline{K}_{B}-2W_2-3W_3)\\
        &+a_3(\frac{2}{3}[Y_{18}]-\frac{1}{3}[Y_5]) \, , \\
         \chi (\two_2)=&\frac{1}{2} C_{\two_2}\, F+a_1 (-\frac{1}{2}[Y_8]+\frac{1}{2}[Y_7]-C_{\two_2}\,  W_2+\frac{1}{2}[Y_1]-\frac{1}{2}[Y_3])+a_4(-\frac{1}{2}[Y_8]+\frac{1}{2}[Y_7])\\
         & +2a_5[Y_4] \, ,  \\
     \chi (\two_1)=&\frac{1}{2}C_{\two_1}\, F+ a_1 (\frac{1}{2}[Y_1]-\frac{1}{2}[Y_3])-a_2[Y_6]-a_3 [Y_5]+a_4(\frac{1}{2} [Y_5]-\frac{1}{2}[Y_6])+2a_5[Y_2] \, , \\
    \chi (\one_{(1,-)})=&-C_{\one_{(1,+)}}\, F+2a_1 [Y_4]+a_2[Y_{15}]+a_3[Y_{15}]+ a_5\,V \, , \\
    \chi (\one_{(1,+)})=&C_{\one_{(1,-)}}\, F-a_1 [Y_1]-a_2[Y_{14}]-a_3[Y_{13}]+ a_5\, V \, , 
    \end{split}
\end{align}
where
\begin{align}
	\begin{split}
    V=&-C_{\one_{(1,+)}}\,  (6\overline{K}_B-2W_2-3W_3)-W_2^2W_3-2W_2W_3^2+2W_2^2\overline{K}_B\\
      &+13W_2W_3{\overline{K}_B} +12W_3^2\overline{K}_B-10W_2\overline{K}_B^2-24W_3\overline{K}_B^2+12\overline{K}_B^3-2W_2W_3S_7\\
      &-4W_3^2S_7-4W_2\overline{K}_BS_7 -W_3\overline{K}_BS_7+8\overline{K}_B^2S_7+2W_2S_7^2+3W_3S_7^2-4\overline{K}_BS_7^2\\
      &+2W_2^2S_9+4W_2W_3S_9-8W_2\overline{K}_BS_9 -9W_3\overline{K}_BS_9+8\overline{K}_B^2S_9+2W_3S_7S_9\\
      & +2W_2S_9^2+3W_3S_9^2-4\overline{K}_BS_9^2 \, .
    \end{split}
\end{align}

\section{Summary of toric tops}
\label{app:B}
In this section we summarize the toric data of all four $SU(2)$ and $SU(3)$ tops over polygon $F_2$, following the prescription of \cite{Bouchard:2003bu}. The factorization of the generic hypersurface is given together with the SR-ideal and the abelian generators.
Furthermore we present the matter loci, the class of fiber component we use for the matter surfaces, and the associated representations.
% \subsection[SU(2) tops over \texorpdfstring{$F_2$}{F2}]{SU(2) tops over \boldmath{$F_2$}}
\begin{figure}[ht]
\begin{center}
\includegraphics[scale=0.5]{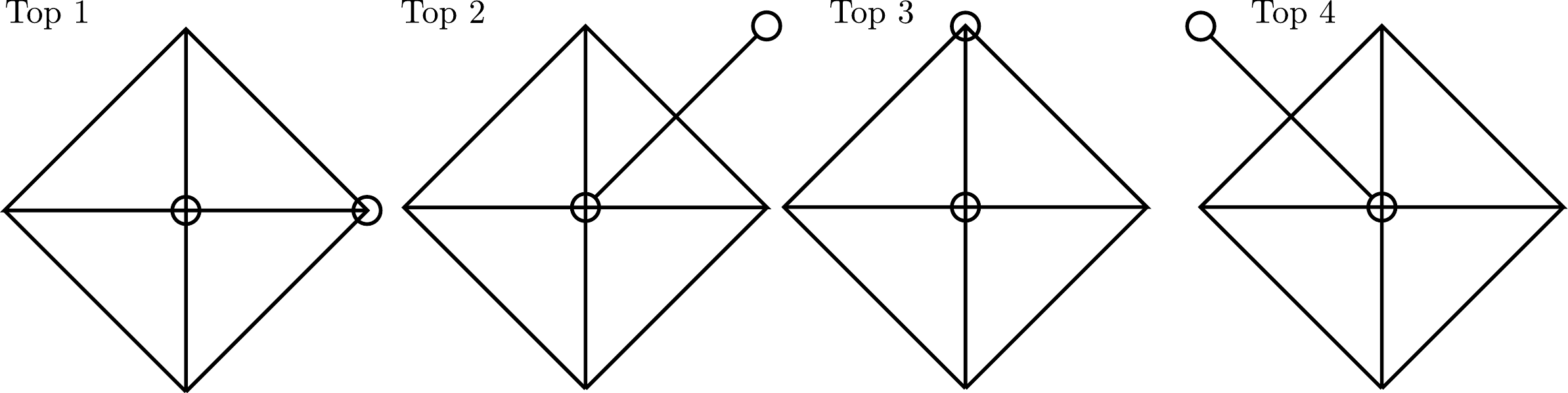}
\caption{\label{fig:su2tops}The toric diagram of the four inequivalent SU(2) tops over $F_2$. }
\end{center}
\end{figure}
 
\begin{table}[ht]
\renewcommand{\arraystretch}{1.3}
\begin{center}
\begin{tabular}{|c|c|c|c|c|}\hline
Factorization:& \multicolumn{4}{|c|}{ $ \begin{array}{lll} 
b_1 \rightarrow e_0 d_1 & b_2 \rightarrow e_0 d_2 &        b_3 \rightarrow e_0 d_3 \\
 b_5\rightarrow d_5 &   b_6 \rightarrow d_6 &  b_7 \rightarrow d_7   \\
  b_8 \rightarrow e_1 d_8 & b_9 \rightarrow e_1d_9  & b_{10} \rightarrow e_1 d_{10}

\end{array}$   }   \\ \hline
SRI:  & \multicolumn{4}{|c|}{$\{ xt, x e_1, y s , t e_0  \} $} \\ \hline
Charge Generators& \multicolumn{4}{|c|}{ $\begin{array}{l} 
D_{U(1)} = [y]-[x] + 1/2 [e_1] \\ 
D_{\bbZ_2} = [x] 
\end{array}$} \\ \hline \hline
Locus & V$(f,g,\Delta)$ & matter $\mathbb{P}^1$ & weight & Rep \\ \hline
$\begin{array}{l}
d_{10}^2 d_5^2 + d_{10} d_6^2 d_8 - 2 d_{10} d_5 d_7 d_8+ d_7^2 d_8^2\\  - 
 d_6 (d_{10} d_5 + d_7 d_8) d_9 + d_5 d_7 d_9^2 = 0
\end{array}$
&  $(0,0,3)$ &$ [e_0][s]$ &$ (-1,1)_{(-\frac12,1)}$ & $\mathbf{2}_{(-\frac12,-)}$ \\ \hline 

$\begin{array}{l}
d_3^2 d_5^2 - d_2 d_3 d_5 d_6 + d_1 d_3 d_6^2 + d_2^2 d_5 d_7\\ - 2 d_1 d_3 d_5 d_7 - 
 d_1 d_2 d_6 d_7 + d_1^2 d_7^2= 0
\end{array}$
&  $(0,0,3)$ &$ [e_1][s]$ &$ (1,-1)_{(-\frac12,0)}$ & $\mathbf{2}_{(\frac12,+)}$ \\ \hline
$4 d_1 d_{10} - d_6^2=0$ & $(1,2,3)$ & - & - &-  \\ \hline  
\end{tabular}
\end{center}
\caption{\label{tab:su2top1}Summary of $SU(2)$ top 1.}
\end{table}

\begin{table}[ht]
\renewcommand{\arraystretch}{0.9}
\begin{center}
\begin{tabular}{|c|c|c|c|c|}\hline
Factorization:& \multicolumn{4}{|c|}{ $ \begin{array}{lll} 
b_1 \rightarrow   d_1 & b_2 \rightarrow e_0 d_2& b_3 \rightarrow e_0^2 d_3  \\
  b_5\rightarrow e_ 1d_5 &  b_6 \rightarrow d_6& b_7 \rightarrow e_0 d_7    \\
   b_8 \rightarrow e_1^2 d_8 &  b_9 \rightarrow e_1d_9& b_{10} \rightarrow   d_{10}

\end{array}$   }   \\ \hline
SRI:  & \multicolumn{4}{|c|}{$\{ xt, y s,x e_0 , s e_1  \} $} \\ \hline
Charge Generators& \multicolumn{4}{|c|}{ $\begin{array}{l} 
D_{U(1)} = [y]-[x] - [e_0] \\ 
D_{\bbZ_2} = [x] 
\end{array}$} \\ \hline \hline
Locus & V$(f,g,\Delta)$ & matter $\mathbb{P}^1$ & Weight & Rep \\ \hline
$d_1 = 0$& $(0,0,3)$& $[e_0][t]$& $(-1,1)_{(1,0)} $ &  $\mathbf{2}_{(1,+)}$ \\ \hline
 $d_{10} = 0 $& $(0,0,3)$& $[e_0][y]$& $(-1,1)_{(-1,1)} $ &  $\mathbf{2}_{(-1,-)}$ \\ \hline

$\begin{array}{l}
-d_{10} d_2^2 + 4 d_1 d_{10} d_3 \\ - d_3 d_6^2 + d_2 d_6 d_7 - d_1 d_7^2=0
\end{array}$
&  $(0,0,3)$ &$ [e_1][y]$ &$ (1,-1)_{(0,0)}$ & $\mathbf{2}_{(0,+)}$ \\ \hline 

$\begin{array}{l}
-d_{10} d_5^2 + 4 d_1 d_{10} d_8 - d_6^2 d_8 \\+ d_5 d_6 d_9 - d_1 d_9^2= 0
\end{array}$
&  $(0,0,3)$ &$ [e_0][e_1 +t+y]$ &$ (1,-1)_{(0,1)}$ & $\mathbf{2}_{(0,-)}$ \\ \hline
$4 d_1 d_{10} - d_6^2=0$ & $(1,2,3)$& - & - &-  \\ \hline 
\end{tabular}
\end{center}
\caption{\label{tab:su2top2}Summary of $SU(2)$ top 2.}
\end{table}

\begin{table}[ht]
\renewcommand{\arraystretch}{0.9}
\begin{center}
\begin{tabular}{|c|c|c|c|c|}\hline
Factorization:& \multicolumn{4}{|c|}{ $ \begin{array}{lll} 
b_1 \rightarrow e_1 d_1 &b_2 \rightarrow d_2& b_3 \rightarrow e_0 d_3  \\
   b_5 \rightarrow e_ 1d_5 &  b_6 \rightarrow d_6&  b_7 \rightarrow e_0 d_7  \\
b_8 \rightarrow e_1 d_8 & b_9 \rightarrow d_9& b_{10} \rightarrow  e_0 d_{10} 

\end{array}$   }   \\ \hline
SRI:  & \multicolumn{4}{|c|}{$\{ xt,   y s ,y e_0, s e_1  \} $} \\ \hline
Charge Generators& \multicolumn{4}{|c|}{ $\begin{array}{l} 
D_{U(1)} = [y]-[x] +1/2 [e_1] \\ 
D_{\bbZ_2} = [x]  + 1/2 [e_1]
\end{array}$} \\ \hline \hline
Locus & V$(f,g,\Delta)$ & matter $\mathbb{P}^1$ & Weight & Rep \\ \hline
 
{\small $\begin{array}{l}
-d_2 d_5 d_6 d_8 + d_1 d_6^2 d_8 + d_2^2 d_8^2 + d_2 d_5^2 d_9 \\ - d_1 d_5 d_6 d_9 - 
 2 d_1 d_2 d_8 d_9 + d_1^2 d_9^2= 0
\end{array}$}
&  $(0,0,3)$ &$ [e_1][x+t]$ &$ (1,-1)_{(\frac12,-\frac12)}$ & $\mathbf{2}_{(-\frac12,\frac12)}$ \\ \hline

{\small $\begin{array}{l}
  d_{10} d_3 d_6^2 - d_{10} d_2 d_6 d_7  - 2 d_{10} d_2 d_3 d_9\\ +d_{10}^2 d_2^2 - 
 d_3 d_6 d_7 d_9 + d_2 d_7^2 d_9 + d_3^2 d_9^2= 0
\end{array}$}
&  $(0,0,3)$ &$ [e_0][x+t]$ &$ (-1,1)_{(\frac12,\frac12)}$ & $\mathbf{2}_{(\frac12,\frac12)}$ \\ \hline

$4 d_2d_{9} - d_6^2=0$ & $(1,2,3)$ & - & - &-  \\ \hline 
\end{tabular}
\end{center}
\caption{\label{tab:su2top3}Summary of $SU(2)$ top 3. In this model the discrete symmetry is enhanced to $\bbZ_2$ due to the form of $D_{\bbZ_2}$. Hence, the representations are labeled by $\bbZ_4$ charges, which here are multiples of $\frac12$ modulo $2\bbZ$.
Note that the charge assignments exhibit the global gauge group structure $[SU(2) \times U(1) \times \bbZ_4]/ ( \bbZ_2^{U(1)} \times \bbZ_2^{{\text{bisec}}})$. Both quotient factors are embedded in the center of $SU(2)$, however, the first one only affects the $U(1)$ charges while the second one restricts the $\bbZ_4$ charges.}
\end{table}
 
\begin{table}[ht]
\renewcommand{\arraystretch}{0.8}
\begin{center}
\begin{tabular}{|c|c|c|c|c|}\hline
Factorization:& \multicolumn{4}{|c|}{ $ \begin{array}{lll} 
b_1 \rightarrow  e_1^2 d_1 & b_2 \rightarrow e_1 d_2& b_3 \rightarrow   d_3   \\
 b_5\rightarrow e_ 1d_5 & b_6 \rightarrow d_6 & b_7 \rightarrow e_0 d_7\\
b_8 \rightarrow   d_8 &  b_9 \rightarrow e_0d_9  & b_{10} \rightarrow  e_0^2 d_{10} 
\end{array}$   }   \\ \hline
SRI:  & \multicolumn{4}{|c|}{$\{ xt, y s,x e_1 , s e_1  \} $} \\ \hline
Charge Generators& \multicolumn{4}{|c|}{ $\begin{array}{l} 
\sigma(s_1) = [y]-[x]  \\ 
\sigma(s^{(2)}) = [x] +[e_1]
\end{array}$} \\ \hline \hline
Locus & V$(f,g,\Delta)$ & matter $\mathbb{P}^1$ & Weight & Rep \\ \hline
$d_3 = 0$& $(0,0,3)$ & $[e_0][y]$& $(-1,1)_{(-1,1)} $ &  $\mathbf{2}_{(-1,-)}$ \\ \hline 
$d_8 = 0$& $(0,0,3)$ & $[e_0][x]$& $(-1,1)_{(1,0)} $ &  $\mathbf{2}_{(1,+)}$ \\ \hline
 
$\begin{array}{l}
d_3 d_5^2 - d_2 d_5 d_6 + d_1 d_6^2 \\+ d_2^2 d_8 - 4 d_1 d_3 d_8=0
\end{array}$
&  $(0,0,3)$ &$ [e_0][x+s]$ &$ (-1,1)_{(0,1)}$ & $\mathbf{2}_{(0,-)}$ \\ \hline 

$\begin{array}{l}
-d_{10} d_6^2 + 4 d_{10} d_3 d_8 - d_7^2 d_8 \\ + d_6 d_7 d_9 - d_3 d_9^2= 0
\end{array}$
&  $(0,0,3)$ &$ [e_1][x+s]$ &$ (1,-1)_{(0,0)}$ & $\mathbf{2}_{(0,+)}$ \\ \hline
$4 d_3 d_{8} - d_6^2=0 $& $(1,2,3)$ & - & - &-  \\ \hline 
\end{tabular}
\end{center}
\caption{\label{tab:su2top4}Summary of $SU(2)$ top 4.}
\end{table}
 
 \clearpage
 %\subsection[SU(3) tops over \texorpdfstring{$F_2$}{F2}]{SU(3) tops over \boldmath{$F_2$}}
\begin{figure}[t]
\begin{center}
\includegraphics[scale=0.5]{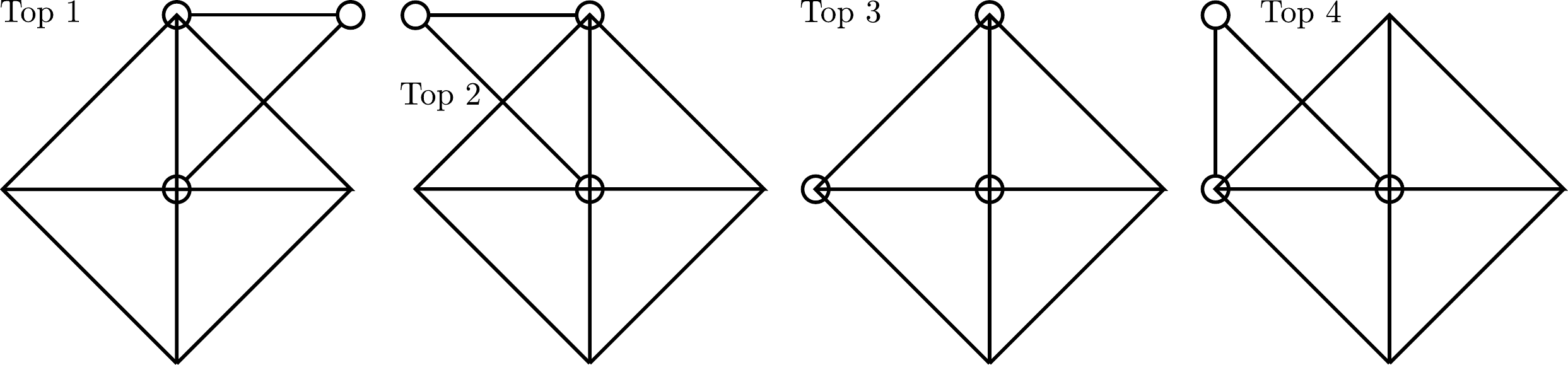}
\caption{\label{fig:su3tops}The toric diagram of the four inequivalent $SU(3)$ tops over $F_2$. }
\end{center}
\end{figure}
\begin{table}
\begin{center}
\begin{tabular}{|c|c|c|c|c|}\hline
Factorization:& \multicolumn{4}{|c|}{{\small $  \begin{array}{lll} 
b_1 \rightarrow  f_2 d_1 &b_2 \rightarrow f_0 f_2d_2 & b_3 \rightarrow  f_0^2 d_3 \\
  b_5\rightarrow f_1 f_2d_5 & b_6 \rightarrow d_6& b_7 \rightarrow f_0 d_7 \\
 b_8 \rightarrow f_1^2  f_2 d_8 & b_9 \rightarrow f_1d_9 & b_{10} \rightarrow  f_0 f_1 d_{10} 
\end{array}$ }  }   \\ \hline
vertices: &  \multicolumn{4}{|c|}{ $f_0:(0,0,1), f_1:(1,1,1), f_2:(0,1,1)$} \\ \hline 
SRI:  & \multicolumn{4}{|c|}{$\{ x t, x f_1, y s, y f_0, t f_2, s f_2, s f_1  \} $} \\ \hline
Charge Generators& \multicolumn{4}{|c|}{ $\begin{array}{l} 
\sigma(s_1) = [y]-[x]+\frac13 (2 [f_1]+[f_2]) \\ 
\sigma(s^{(2)}) = [x] ++\frac13 (2 [f_1]+[f_2])
\end{array}$} \\ \hline \hline
Locus & V$(f,g,\Delta)$ & matter $\mathbb{P}^1$ & Weight & Rep \\ \hline
$d_1 = 0$& $(0,0,4)$ & $[f_0][t]$& $(1,0)_{(\frac23,\frac13)} $ &  $\mathbf{3}_{(\frac23, -)}$ \\ \hline

$d_{10} d_6 - d_7 d_9 = 0$& $(0,0,4)$ & $[f_2][x]$& $(0,-1)_{(\frac23,-\frac23)} $ &  $\mathbf{3}_{(\frac23, +)}$ \\ \hline
 
$\begin{array}{l}
d_3 d_6^2 - d_2 d_6 d_7 \\ + d_1 d_7^2=0
\end{array}$
&  $(0,0,4)$ &$ [f_1][y]$ &$ (-1,1)_{(-\frac13, \frac13)}$ & $\mathbf{3}_{(-\frac13, -)}$ \\ \hline 

$\begin{array}{l}
d_6^2 d_8 - d_5 d_6 d_9 \\ + d_1 d_9^2 = 0
\end{array}$
&  $(0,0,4)$ &$ [f_0][x]$ &$ (0,1)_{(\frac13, -\frac23)}$ & $\mathbf{3}_{(-\frac13,+)}$ \\ \hline
$d_6=0 $ & $(2,2,4)$ & - & - &-  \\ \hline 
\end{tabular}
\end{center}
\caption{\label{tab:su3top1}Summary of $SU(3)$ top 1.}
\end{table}

\begin{table}[ht]
\begin{center}
\begin{tabular}{|c|c|c|c|c|}\hline
Factorization:& \multicolumn{4}{|c|}{ {\small $  \begin{array}{lll} 
b_1 \rightarrow  f_2 d_1 & b_2 \rightarrow f_1d_2& b_3 \rightarrow  f_0 f_1 d_3 \\
  b_5\rightarrow f_1 f_2d_5 & b_6 \rightarrow d_6 & b_7 \rightarrow f_0 d_7 \\
 b_8 \rightarrow  f_2 d_8 & b_9 \rightarrow f_0 f_2 d_9& b_{10} \rightarrow  f_0^2 f_2 d_{10} 
\end{array}$ }  }   \\ \hline
vertices: &  \multicolumn{4}{|c|}{ $f_0:(0,0,1), f_1:(-1,1,1), f_2:(0,1,1)$} \\ \hline 
SRI:  & \multicolumn{4}{|c|}{$\{ x t, x f_2, y s, y f_0, t f_1, s f_1, s f_2  \} $} \\ \hline
Charge Generators& \multicolumn{4}{|c|}{ $\begin{array}{l} 
\sigma(s_1) = [y]-[x]-\frac13 ([f_1]-[f_2]) \\ 
\sigma(s^{(2)}) = [x] +\frac23 (2 [f_1]+[f_2])
\end{array}$} \\ \hline \hline
Locus & V$(f,g,\Delta)$ & matter $\mathbb{P}^1$ & Weight & Rep \\ \hline
$d_8 = 0$& $(0,0,4)$ & $[f_0][x]$& $(1,0)_{(\frac23, \frac13)} $ &  $\mathbf{3}_{(\frac23,-)}$ \\ \hline

$-d_3 d_6 + d_2 d_7 = 0$& $(0,0,4)$ & $[f_2][t]$& $(0,-1)_{(\frac23,-\frac23)} $ &  $\mathbf{3}_{(\frac23,+)}$ \\ \hline
 
$\begin{array}{l}
-d_2 d_5 d_6 + d_1 d_6^2\\ + d_2^2 d_8 =0
\end{array}$
&  $(0,0,4)$ &$ [f_0][t]$ &$ (0,1)_{(\frac13, \frac23)}$ & $\mathbf{3}_{(-\frac13, +)}$ \\ \hline 

$\begin{array}{l}
d_10 d_6^2 + d_7^2 d_8 \\ - d_6 d_7 d_9 = 0
\end{array}$
&  $(0,0,4)$ &$ [f_1][y]$ &$ (-1,1)_{(-\frac13, \frac13)}$ & $\mathbf{3}_{(-\frac13, -)}$ \\ \hline
$d_6=0$ &$ (2,2,4)$ & - & - &-  \\ \hline 
\end{tabular}
\end{center}
\caption{\label{tab:su3top2}Summary of $SU(3)$ top 2.}
\end{table} 

\begin{table}[htt]
\begin{center}
\begin{tabular}{|c|c|c|c|c|}\hline
Factorization:& \multicolumn{4}{|c|}{ {\small $  \begin{array}{lll} 
b_1 \rightarrow f_0 f_2 d_1 & b_2 \rightarrow f_0d_2 & b_3 \rightarrow  f_0^2 f_1 d_3  \\
 b_5\rightarrow  f_2d_5 &b_6 \rightarrow d_6 & b_7 \rightarrow f_0 f_1 d_7  \\
 b_8 \rightarrow  f_1 f_2^2 d_8 &  b_9 \rightarrow f_1 f_2 d_9 & b_{10} \rightarrow  f_1 d_{10}
\end{array}$ }  }   \\ \hline
vertices: &  \multicolumn{4}{|c|}{ $f_0:(0,0,1), f_1:(0,1,1), f_2:(1,1,1)$} \\ \hline 
SRI:  & \multicolumn{4}{|c|}{$\{ x t, x f_1, x f_2, y s, y f_1, t f_0, s f_2  \} $} \\ \hline
Charge Generators& \multicolumn{4}{|c|}{ $\begin{array}{l} 
\sigma(s_1) = [y]-[x]+\frac23 ([f_1]+2[f_2]) \\ 
\sigma(s^{(2)}) = [x]
\end{array}$} \\ \hline \hline
Locus & V$(f,g,\Delta)$ & matter $\mathbb{P}^1$ & Weight & Rep \\ \hline
$d_{10} = 0$& $(0,0,4)$ & $[f_0][y]$& $(0,1)_{(-\frac23,1)} $ &  $\mathbf{3}_{(\frac23,- )}$ \\ \hline

$d_2 d_5 - d_1 d_6 = 0$& $(0,0,4)$ & $[f_1][t]$& $(-1,1)_{(\frac23,0)} $ &  $\mathbf{3}_{(\frac23, +)}$ \\ \hline
 
$\begin{array}{l}
d_10 d_2^2 + d_3 d_6^2 \\ - d_2 d_6 d_7= 0
\end{array}$
&  $(0,0,4)$ &$ [f_2][t]$ &$ (1,-1)_{( \frac13,0)}$ & $\mathbf{3}_{(\frac13, +)}$ \\ \hline 

$\begin{array}{l}
d_10 d_5^2 + d_6^2 d_8 \\ - d_5 d_6 d_9 = 0
\end{array}$
&  $(0,0,4)$ &$ [f_0][s]$ &$ (1,0)_{(-\frac13,1)}$ & $\mathbf{3}_{(-\frac13,-)}$ \\ \hline
$d_6 =0$ & $(2,2,4)$ & - & - &-  \\ \hline 
\end{tabular}
\end{center}
\caption{\label{tab:su3top3}Summary of $SU(3)$ top 3. }
\end{table}

\begin{table}[ht]
\begin{center}
\begin{tabular}{|c|c|c|c|c|}\hline
Factorization:& \multicolumn{4}{|c|}{ {\small $  \begin{array}{lll} 
b_1 \rightarrow f_0  f_1^2 d_1 & b_2 \rightarrow f_0 f_1d_2 & b_3 \rightarrow  f_0  d_3   \\
  b_5\rightarrow f_1  d_5 &  b_6 \rightarrow d_6 & b_7 \rightarrow f_0 f_2 d_7 \\
b_8 \rightarrow  f_1 f_2 d_8 & b_9 \rightarrow  f_2 d_9 & b_{10} \rightarrow  f_0 f_2^2 d_{10} 
\end{array}$ }  }   \\ \hline
vertices: &  \multicolumn{4}{|c|}{ $f_0:(0,0,1), f_1:(0,1,1), f_2:(1,0,1)$} \\ \hline 
SRI:  & \multicolumn{4}{|c|}{$\{ x t, x f_2, y s, y f_0, y f_2, s f_1, t f_0  \} $} \\ \hline
Charge Generators& \multicolumn{4}{|c|}{ $\begin{array}{l} 
\sigma(s_1) = [y]-[x]+\frac13 (2 [f_1]+[f_2]) \\ 
\sigma(s^{(2)}) = [x]+\frac13 (2 [f_1]+[f_2])
\end{array}$} \\ \hline \hline
Locus & V$(f,g,\Delta)$ & matter $\mathbb{P}^1$ & Weight& Rep \\ \hline
$d_{3} = 0$& (0,0,4)& $[f_1][t]$& $(-1,1)_{(\frac23,-\frac13)} $ &  $\mathbf{3}_{(\frac23, -)}$ \\ \hline

$-d_6 d_8 + d_5 d_9 = 0$& (0,0,4)& $[f_0][x]$& $(1,0)_{(\frac23,\frac23)} $ &  $\mathbf{3}_{(\frac23, +)}$ \\ \hline
 
$\begin{array}{l}
d_3 d_5^2 - d_2 d_5 d_6 \\ + d_1 d_6^2 = 0
\end{array}$
&  $(0,0,4)$ &$ [f_2][s]$ &$ (0,-1)_{(-\frac13,-\frac13)}$ & $\mathbf{3}_{(-\frac13, -)}$ \\ \hline 

$\begin{array}{l}
d_{10} d_6^2 - d_6 d_7 d_9 \\ + 
   d_3 d_9 = 0
\end{array}$
&  $(0,0,4)$ &$ [f_1][x]$ &$ (-1,0)_{(\frac13,-\frac23)}$ & $\mathbf{3}_{(-\frac13,+)}$ \\ \hline
$d_6=0 $& $(2,2,4)$ & - & - &-  \\ \hline 
\end{tabular}
\end{center}
\caption{\label{tab:su3top4}Summary of $SU(3)$ top 4. }
\end{table} 
\clearpage

\bibliography{Z2-SM}{}
\bibliographystyle{JHEP}

\end{document}